\documentclass[preprint2]{aastex631}
\usepackage{longtable}
\usepackage{amsmath}
\usepackage{todonotes}
\usepackage{subfigure}
\usepackage{multirow}

\shorttitle{}
\shortauthors{Dornan \& Harris}

\graphicspath{{./}{figures/}}

\begin{document}



\title{Major Mergers Mean Major Offset: Drivers of Intrinsic Scatter in The $M_{GCS} -M_h$ Scaling Relation for Massive Elliptical Galaxies}


\author[0000-0002-7731-1291]{Veronika Dornan}
\affiliation{Department of Physics and Astronomy \\
McMaster University \\
Hamilton, ON, L8S 4M1}

\author[0000-0001-8762-5772]{William E. Harris}
\affiliation{Department of Physics and Astronomy \\
McMaster University \\
Hamilton, ON, L8S 4M1}

\begin{abstract}

In this work we determine the total globular cluster (GC) counts and globular cluster system (GCS) total mass estimates for 27 extremely massive elliptical galaxies. The GC 2D spatial distributions of these galaxies were created from photometry of HST images using DOLPHOT in the near-IR wavelength range. The projected radial density profiles of these GCSs were determined using a Voronoi tessellation-based technique introduced in our previous paper. We then plot these galaxies on the GCS - halo mass relation alongside previously studied galaxies in the literature.  The relation now extends across seven decades of halo mass. We find that the 1:1 slope of this relation holds out to the highest mass galaxies, although extremely massive BCG galaxies are shifted to higher GCS masses than their lower-mass galaxy counterparts. We find a negative correlation with massive galaxies' offset from the GCS - halo mass relation and the steepness of their GCS density profiles, and that this is being driven by the red GC populations. We suggest that the biggest influence in intrinsic scatter in the GCS - halo mass relation for massive galaxies is through a few major mergers resulting in accretion of massive satellites with old, red GC populations, rather than many accretions of small satellites with younger, blue GC populations.
\end{abstract}


\section{Introduction}\label{sec:intro}

Globular clusters (GCs) have long been a useful tool in tracing galaxy structure and evolution. These massive stellar clusters, hosting as many as $10^7$ stars, are some of the oldest surviving objects within galaxies across all masses \citep{Vandenberg13, Beasley20}. Their compact size and strong gravitational boundedness also ensure that these clusters can survive tidal disruption and merger events that their host galaxies have undergone \citep{Marta23, Joschko24}. As a result, the properties of a globular cluster system (GCS) can correlate with its host galaxy's global properties and merger history \citep{Belokurov23,Belokurov24, Newton24, Federle24, Mirabile24}. 

One scaling relation of particular interest is that between the total number of GCs ($N_{GC}$), or total mass enclosed in a galaxy's GCS ($M_{GCS}$) and the total mass of the galaxy, typically dominated by its dark matter halo ($M_h$) \citep{Blakeslee97}. This relation has been known and studied both through observations and simulations for decades. It has been found that across a range of at least $10^6$ in galaxy mass this relation holds a tight 1:1 linear shape \citep{Spitler09, Hudson14,Harris15, Burkert20, Dornan23}. It is well known that this relation, particularly for massive galaxies, is driven by mergers, as they can grow both the dark matter halos and GCSs of galaxies \citep{Choksi19, Chen23, Valenzuela24}. However, there exists uncertainty over what drives the variation in GCS mass for galaxies of the same halo mass that is observed.

Observationally, there is a significant amount of GCS data for galaxies with halo masses in the range of $10^{11}-10^{13}M_{\odot}$ \citep{Peng08,Villegas10,Lim24}, becoming more sparse above this mass range, where brightest cluster galaxies (BCGs) dominate. For the data that is available in the literature, the methods used to determine both the $M_{GCS}$ and $M_h$ estimates are not always consistent. This makes it difficult to get an accurate observational understanding of the behaviour of this relation across different mass ranges. 

The issue of consistency is especially interesting for galaxies with halo masses above $10^{13} M_{\odot}$, which are mostly BCGs (brightest cluster galaxies) with extended halos that may not be enclosed in the imaging used to determine GC counts. They also typically reside in high density clustered environments, which can also introduce difficulty in determining the exact boundary of the galaxy's GCS and the contribution from the host galaxy cluster's intracluster medium, which hosts its own GC population.

In these clustered environments BCGs typically have multiple nearby satellite galaxies that are included in imaging of the target BCG. The GCSs of these satellite galaxies also add difficulty in determining the GC counts for the target BCG, as these satellites host their own GCSs. It can sometimes become unclear which satellite galaxies should be masked from an image to ensure GC contamination is minimized, and which should be left in to ensure GC counts for the target BCG are maximized.

All of these considerations added together make studying GCSs around extremely massive BCGs not as straightforward as their lower-mass counterparts, but the process can be optimized through consistent methodology. In this paper we apply a new Voronoi tessellation-based method of determining GC radial density profiles to 27 massive ellipticals, to replace the standard annulus-based method. 

This new Voronoi method was introduced in \cite{Dornan24}, hereafter referred to as Paper I, and it was found that when applied to circularly symmetric single systems of approximately 2000 or more detected objects the Voronoi method outperformed the conventional annulus method for both accuracy and precision of fits to the radial density profiles. In this paper we now will apply this method to a sample of more complicated, observed systems, many of which host significant satellite systems, have extended halos, or are highly elliptical. We derive more accurate, precise, and methodologically consistent GCS mass estimates for a large sample of galaxies with halo masses above $10^{13} M_{\odot}$. This will allow for the high-mass end of the $M_{GCS} - M_h$ relation to be far more accurately constrained observationally than previous mass estimates have allowed.

\begin{center}
\begin{table*}[h!tb]
    \centering
    \caption{List of target galaxies} \label{tab:Names}
    \begin{tabular}{ccccccccc}
    \hline \hline
    Target Name & $l$ & $b$ & $A_I$ & $(m-M)_I$ & $M_V^T$ & $M_K$ & $D(Mpc)$ & BCG/NMCG \\
    (1) & (2) & (3) & (4) & (5) & (6) & (7) \\
    \hline
    J13481399-3322547 & 316.35\textdegree & +28.01\textdegree & 0.082 & $36.335 \pm 0.093$ & -21.7 & -25.5 & $178 \pm 8$ & BCG  \\
    J13280261-3145207 & 311.96\textdegree & +30.47\textdegree & 0.079 & $36.446 \pm 0.093$ & -22.0 & -25.3 & $187\pm 8$ & NMCG \\
    J13275493-3132187 & 311.97\textdegree & +30.69\textdegree & 0.076 & $36.839 \pm 0.093$ & -23.3 & -26.1 &  $225 \pm 10$ & NMCG \\
    J13272961-3123237 & 311.89\textdegree & +30.85\textdegree & 0.088 & $36.679 \pm 0.093$ & -23.3 & -26.2 & $208\pm9$ & NMCG \\
    ESO 509-G067 & 314.69\textdegree & +34.75\textdegree & 0.103 & $36.023 \pm 0.094$ & -23.3 & -25.8 & $153 \pm 7$ & NMCG \\
    ESO 509-G020 & 312.83\textdegree & +34.81\textdegree & 0.086 & $35.957 \pm 0.094$ & -23.3 & -25.6 & $149 \pm 6$ & NMCG \\
    ESO 509-G008 & 312.47\textdegree & +34.78\textdegree & 0.080 & $36.031 \pm 0.093$ & -23.0 & -26.1 & $155\pm7$ & BCG \\
    ESO 444-G046 & 311.99\textdegree & +30.73\textdegree & 0.076 & $36.635 \pm 0.093$ & -24.8 & -27.1 &  $205\pm 9$ & BCG \\
    ESO 383-G076 & 316.32\textdegree & +28.55\textdegree & 0.083 & $36.223 \pm 0.093$ & -24.2 & -26.8 &  $169\pm7$ & BCG \\
    ESO 325-G016 & 314.72\textdegree & +23.64\textdegree & 0.123 & $36.214 \pm 0.093$ & -22.3 & -25.4 & $165\pm7$ & BCG \\
    ESO 325-G004 & 314.08\textdegree & +23.57\textdegree & 0.092 & $35.958 \pm 0.093$ & -23.3 & -26.2 & $149\pm6$ & BCG \\
    ESO 306-G017 & 246.41\textdegree & -30.29\textdegree & 0.041 & $36.069 \pm 0.094$ & -24.3  & -26.5 & $161\pm7$ & BCG \\
    NGC 1129 & 146.34\textdegree & -15.63\textdegree & 0.279 & $34.541 \pm 0.108$ & -22.9 & -26.1 & $71 \pm3$ & BCG \\
    NGC 1132 & 176.45\textdegree & -51.07\textdegree & 0.080 & $34.993 \pm 0.094$ & -22.5 & -25.7 & $96\pm4$ & NMCG \\
    NGC 1272 & 150.52\textdegree & -13.32\textdegree & 0.245 & $34.512 \pm 0.097$ & -23.3 & -25.6 & $74\pm 3$ & NMCG \\
    NGC 1278 & 150.56\textdegree & -13.21\textdegree & 0.251 & $34.518 \pm 0.097$ & -22.3 & -25.2 & $71\pm3$ & NMCG \\
    NGC 3258 & 272.90\textdegree & +18.82\textdegree & 0.123 & $33.457 \pm 0.095$ & -22.1 & -25.1 &  $46\pm2$ & BCG \\
    NGC 3268 & 272.94\textdegree & +19.18\textdegree & 0.157 & $33.491 \pm 0.096$ & -23.2 & -25.2 & $46\pm2$ & BCG \\
    NGC 3348 & 134.63\textdegree & +41.35\textdegree & 0.113 & $33.125 \pm 0.094$ & -21.7& -25.1 & $40\pm2$ & NMCG \\
    NGC 4696 & 302.40\textdegree & +21.56\textdegree & 0.170 & $33.604 \pm 0.099$ & -24.2 & -26.3 &  $49\pm2$ & BCG \\
    NGC 4874 & 58.08\textdegree & +88.01\textdegree & 0.014 & $35.094 \pm 0.094$ & -23.7 & -26.2 & $104\pm4$ & BCG \\
    NGC 4889 & 57.19\textdegree & +87.89\textdegree & 0.015 & $35.095 \pm 0.094$ & -23.8 & -26.7 & $104\pm4$ & NMCG \\
    NGC 6166 & 62.93\textdegree & +43.69\textdegree & 0.017 & $35.611 \pm 0.094$ & -23.6 & -26.43 & $131\pm6$ & BCG \\
    NGC 7626 & 87.86\textdegree & -48.38\textdegree & 0.110 & $33.616 \pm 0.096$ & -22.4 & -25.5 & $50\pm2$ & BCG \\
    NGC 7720 & 103.50\textdegree & -33.07\textdegree & 0.108 & $35.669 \pm 0.096$ & -23.2 & -26.2 & $124\pm5$ & BCG \\
    UGC 10143 & 28.91\textdegree & +44.52\textdegree & 0.048 & $36.069 \pm 0.095$ & -24.4 & -25.8 &  $160\pm7$ & BCG \\
    UGC 9799 & 9.42\textdegree & +50.12\textdegree & 0.057 & $35.936 \pm 0.094$ & -23.6 & -26.3 & $150\pm6$ & BCG \\
    \hline
    \end{tabular}
\item{} \footnotesize{\textit{Key to columns:} (1) Galaxy identification; (2,3) Galactic longitude and latitude; (4) foreground extinction; (5) apparent distance modulus; (6) total visual absolute magnitude; (7) total K-band absolute magnitude;  (8) adopted distance in Mpc; (9) BCG or NMCG classification.  All images are taken with the HST ACS/WFC camera.}
\end{table*}
\end{center}

\section{Methods}\label{sec:methods}

\subsection{Galaxy Sample}\label{sec:sample}

The sample of galaxies analyzed in this work consists of 27 massive elliptical galaxies, most of which are classified as BCGs, and the remainder would be classified as the second or third brightest galaxies in their respective clusters, which we refer to here as next massive cluster galaxies (NMGCs). The first 11 galaxies listed in Table \ref{tab:Names} were previously studied in \cite{Dornan23} using an annulus-based method to determine their GC radial density profiles, where here we will be updating their profile fits using the Voronoi method. The GCSs of the last 16 galaxies listed in Table \ref{tab:Names} were studied photometrically in \cite{Harris23}, although their GC radial density profiles were not determined. 

Table \ref{tab:Names} lists all galaxies in the sample as well as their Galactic latitudes and longitudes, their extinctions, distance moduli, total visual absolute magnitudes, total K-band absolute magnitudes, distances in Megaparsecs, and their classification as a BCG or NMCG, taken from the NASA/IPAC Extragalactic Database. The distances were determined using $H_0 = (70 \pm 3 km/s/Mpc)$. Although some the galaxies in our sample have updated surface brightness fluctuation distances available, we have decided to use the same distance calculation for all galaxies for internal consistency. We have also chosen to use $H_0 = (70 \pm 3 km/s/Mpc)$ for consistency with the analysis done in \cite{Dornan23} and \cite{Harris23}, but included an uncertainty which covers the $H_0$ values found by \cite{wmap13}, \cite{planck15}, and \cite{planck18}. This sample of galaxies was selected from the HST archive based on their distances, filters used for imaging, and depth of exposure times. These selection criteria ensured that the GCs hosted by these galaxies would appear star-like and have high signal-to-noise ratios (SNRs), making them easily identifiable by the photometry code used \citep{Dornan23, Harris23}

\subsection{Photometry}\label{sec:phot}

Photometry for all galaxies in this sample was carried out using DOLPHOT \citep{Dolphin00}, a package originally designed for stellar photometry, but which is widely used for GC photometry at distances above $\sim 25$Mpc, as at these distances GCs appear morphologically star-like. 

All photometric data used in this paper is from \cite{Dornan23} and \cite{Harris23}. Key photometric parameters used can be found in Table 1 of \cite{Harris23}. Once DOLPHOT identified star-like objects in each image, those objects were culled to leave behind only GC candidates. For the galaxies studied in \cite{Dornan23} this was done purely through chi, sharp, and signal-to-noise ratios, while for the galaxies studied in \cite{Harris23}, due to the multiple filters available, star-like objects could also be culled using colour-magnitude diagrams. The limiting magnitudes for each image were determined by adding artificial GCs with a variety of magnitudes to each image, re-running the photometric process, and determined at which magnitude less than 50\% of the artificial stars were identified by DOLPHOT. These limiting magnitudes can be found in Table 2 of \cite{Dornan23} and Table 3 of \cite{Harris23}.

\subsection{Determining Radial Profiles with Voronoi Tessellations}\label{sec:profiles}

Once a 2D distribution of all detected and limiting-magnitude-corrected GCs around each target galaxy has been created, we then move on to create GC radial density profiles for each galaxy. For some of our sample these profiles have already been determined in \cite{Dornan23} with a standard circular annulus method. However, our aim in this work is to improve the fits of those profiles with a new Voronoi method, as was described in Paper I. 

In brief, the Voronoi method begins by taking the 2D distribution of objects and creating a Voronoi tessellation map, where each object is contained in a polygon whose area is inversely proportional to its local density. The Voronoi tessellations were computed using the built-in Voronoi function in the scipy.spatial package \citep{scipy}. The areas of these tessellations are then inverted to obtain a density value for each detected object. As described in Paper I, in order to reduce stochastic scatter for the radial profile, the individual tessellations are then spatially binned into groups of five and combined into one larger cell with an averaged radius value and combined density value. The densities of these binned cells are then plotted as a function of radius. In order to further reduce stochastic scatter, the mean density of the cells in each interval of radius is determined and any cells with densities greater than 1.5 standard deviations from the mean are culled. 

One of the benefits of this Voronoi method over the annulus method that was not explored in Paper I was the ability to better identify satellite galaxies with contaminating GCSs. Although satellite galaxies can be identified visually in the GC spatial distribution, it is not always immediately clear which galaxies are large and extended enough to warrant removal. Typically, one method of identifying significant satellite systems is to find GC overdensities in the radial and azimuthal profiles and mask out the satellite that would correspond to the overdensity's position and re-run the whole process. One of the difficulties with the annulus method was that it was very difficult to accurately identify overdensities when only $\sim 20-25$ density values are available and can be subject to their own stochastic scatter.

\begin{figure*}[h!tb]
    \centering
    \includegraphics[width=0.9\textwidth]{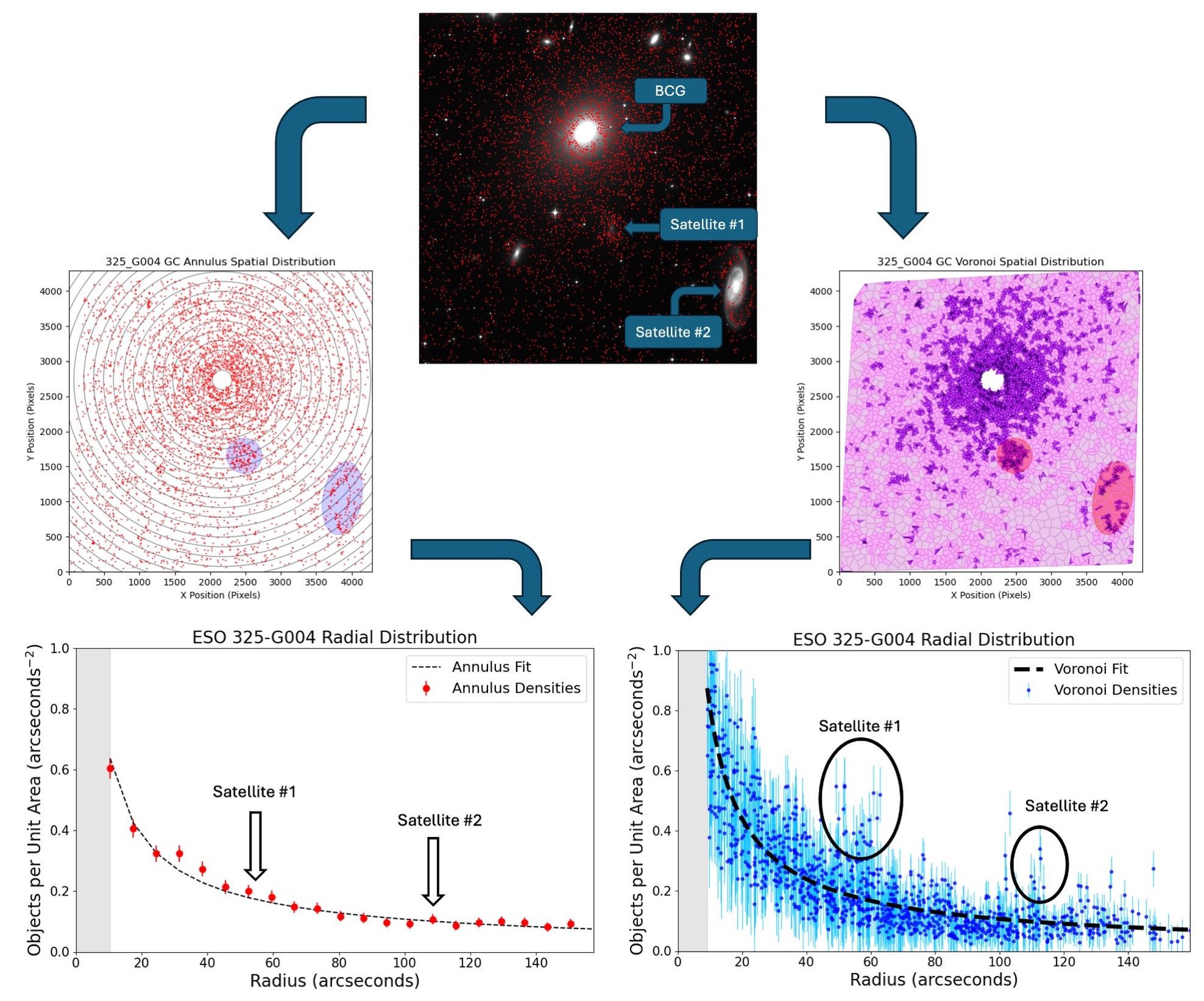}
    \caption{\label{fig:removal} \footnotesize{An example of the spatial distributions and radial density profiles of ESO 325-G004 using both the annulus and Voronoi methods, before removal of any satellites. On the left it is difficult to identify the satellites and determine if they are affecting the fit to the data. On the right it is much more clear where the satellites are, how extended they are, and by how much they are driving up the average density at their radii.}}
\end{figure*}

However with the Voronoi method, the increased number of density values available makes it much more apparent which satellite systems need to be removed and which could remain without significantly impacting the profile fit. As a result, many of the satellites that were removed from the galaxies studied in \cite{Dornan23} (see figure 4) were found unnecessary to mask, and instead remained in this work. In the case of ESO 325-G004, there was one satellite that was not removed in \cite{Dornan23} that was now identified to be significant and removed. An example of this can be see in Figure \ref{fig:removal}.

\begin{figure*}[h!tb]
    \centering
    \includegraphics[width=0.93\textwidth]{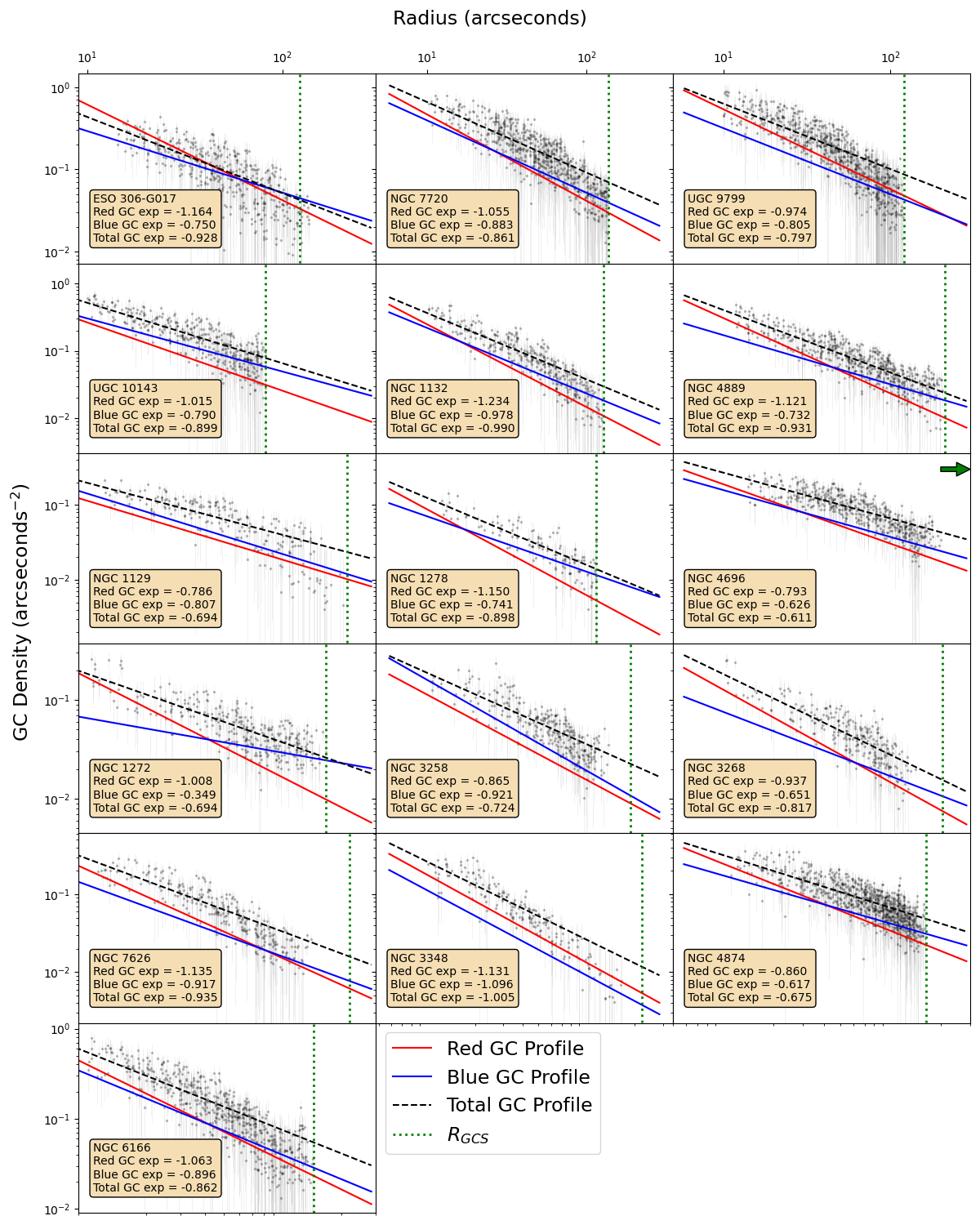}
    \caption{\label{fig:grid_colour} \footnotesize{The red, blue and total GC density profiles for the 16 galaxies in this sample with colour information available, plotted in log-log space. The green vertical dashed lines represent the radius of the GCS and outer bound of integration for determining each galaxy's $N_{GC}$, defined here as $0.1R_{vir}$. Each galaxy's exact $R_{GCS}$ values can be found in Table \ref{tab:systems}. The names of each galaxy and the exponents of their powerlaw fits for the total, red, and blue GCS profiles are shown in the inset boxes.}}
\end{figure*}

\begin{figure*}[h!tb]
    \centering
    \includegraphics[width=0.97\textwidth]{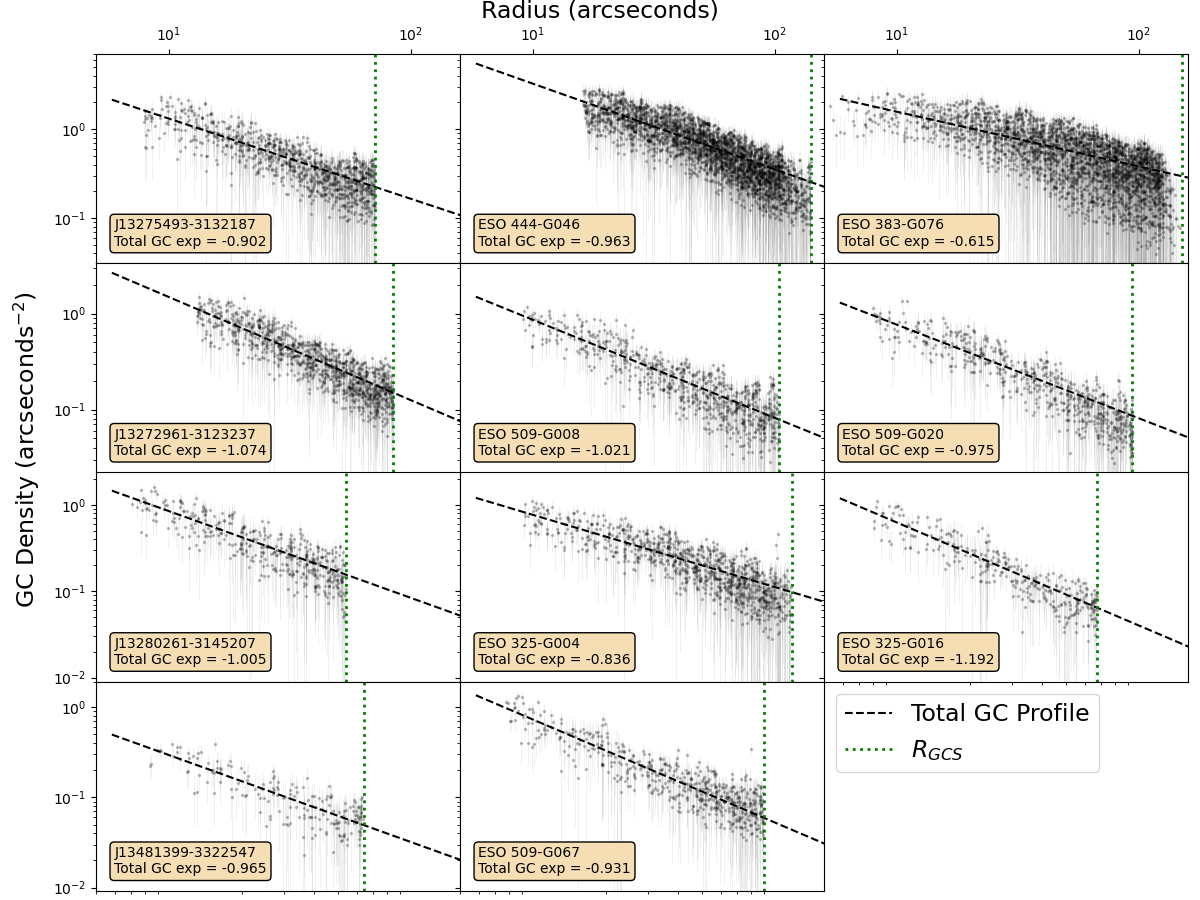}
    \caption{\label{fig:grid_no_colour} \footnotesize{Same as for Figure \ref{fig:grid_colour}, but for the 11 galaxies in this sample with no colour information.}}
\end{figure*}

Once this culling and satellite system removal is complete, a powerlaw is fit to the remaining data using a boostrapping technique with 1000 iterations. An example of this fit with the Voronoi datapoints can be seen in Figure \ref{fig:removal} for ESO 325-G004. The fits for all the galaxies in the sample with colour information are shown in Figure \ref{fig:grid_colour}, and the fits for all the galaxies in the sample without colour information are shown in Figure \ref{fig:grid_no_colour}. As was discussed in the results of Paper I, the Voronoi method returns radial profile fits with lower uncertainties, subsequently allowing for more precise estimates of total GC counts.    

\subsection{Determining GCS Masses}\label{sec:GCS}

Once the radial density profiles of the galaxies were determined they were then integrated out to the defined radius of the GCS. Here we adopt the same definition of GCS radius as in \cite{Dornan23}, where $R_{GCS} = 0.1 R_{vir}$. This is to ensure that we have a standardized definition of GCS size in order to effectively compare the resulting GCS masses of these sample galaxies to one another. We define the $R_{GCS}$ using Equation \ref{eq:vir_rad}, below.
\begin{align}\label{eq:vir_rad}
R_{GCS} &\equiv 0.1 R_{vir} \nonumber\\
 &= 0.1 \Big[\frac{3 M_{vir}}{4 \pi \cdot 200 \rho_c}\Big]^{1/3}\\
 &= 0.1 \Big[\frac{G M_{vir}}{100 H_0^2}\Big]^{1/3} \nonumber
\end{align}
where $M_{vir}$ is in Solar masses and $H_0$ in km s$^{-1}$ Mpc$^{-1}$.  Here $\rho_c = 3 H_0^2/8 \pi G$ is the cosmological critical density and it is assumed for the purposes of this study that $M_{vir} \simeq M_h$. We note that this definition does mean that the $R_{GCS}$ becomes dependent on the $M_h$ for each galaxy, but does so on the cube root, making it not strongly sensitive to small variations in $M_h$. Although it \textit{is} sensitive to large variations in $M_h$, it is still to a smaller extent than by which GCS size in general scales with $M_h$ \citep{Forbes17, Hudson18}.

Due to the high background light intensity at the centres of these galaxies DOLPHOT is unable to identify any GCs in the innermost radii of the galaxies in the sample. Since we have a lack of information in regards to the radial density profiles at these radii we simply assume a constant density here, equivalent to the density of the innermost data available. This approximation is reasonable as the area of this region is small and observations of the Milky Way and M31, for which we do have GC data at small radii, find that the GC density does in fact level off towards the bulge \citep{Huxor11}. Other studies of massive ellipticals also find that the GC radial density profiles begin to flatten before detection becomes no longer possible \citep{Capuzzo09}. It should be noted that for BCGs this flattening of the profile can typically occur between $2-10$ kpc \citep{Peng11}, and as can be seen in column 3 of Table \ref{tab:systems}, the inner limits of our photometry for some of our galaxies is within or exceeds this range. We caution that depending on assumptions made about where these GC density profiles level off it is possible for the $N_{GC}$ estimates of this sample to increase by approximately 2\%.

Thus, the total number of globular clusters in the image can be defined by Equation \ref{eq:integral}, as shown below.
\begin{equation}\label{eq:integral}
    N_{GC} = \int_0^{R_{in}} \sigma_{in} 2 \pi r dr + \int_{R_{in}}^{R_{GCS}} 2 \pi r \sigma_{cl} dr
\end{equation}
Here $R_{in}$ represents the inner limit of the GC photometry, $R_{GCS}$ represents the radial size of the GCS, $\sigma_{in}$ represents the adopted, constant GC density in the inner region, and $\sigma_{cl}$ represents the variable density of GCs in arcseconds$^{-1}$ in at all other radii.

Although we have already removed background GCs and attempted to minimize the contamination from the ICM by using a standard $R_{GCS}$ definition, within our image we likely still have a fraction of our detected GCs which are actually ICGCs. Due to the small field of view of our HST images it is outside the scope of this work to determine the ICGC background density within all the different environments which host the various galaxies in our sample. Instead, we assume that $3\% \pm 3\%$ of the final, integrated $N_{GC}$ estimate are from the ICM and are thus subtracted from our totals. This estimate would cover very low-density environments where the ICGC contamination is estimated to be negligible \citep{NGVS20, Harris20}, as well as richer environments where the contamination has been estimated to be as much as $5\% -6\%$ \citep{Madrid18, Harris23}.

Finally, this estimate must be corrected to account for the GCs that have magnitudes dimmer than the limiting magnitude of the image. For the galaxies in our sample we assume that their GC luminosity functions (GCLFs) take on a well-defined, classical log-normal shape, with peaks at $M_I = -9.0 \pm 0.3$ and Gaussian dispersions of $\sigma_g = 1.30$ \citep{Harris14}. The fraction of undetected GCs in each image can then be calculated through comparing the peak GCLF magnitude to the absolute limiting magnitude of the images of each galaxy in the sample.

With the final, corrected estimate of the total GC count for each galaxy, which, as can be seen in Table \ref{tab:systems}, is on the order of $10^4 - 10^5$, we can then convert this to total GCS mass by simply multiplying it by the average single GC mass for each galaxy. For massive ellipticals such as these, there exists a well-defined, shallow relation between average GC mass and host galaxy luminosity \citep{Villegas10,Harris17}, as described in Equation \ref{eq:avgGC} below.
\begin{equation}\label{eq:avgGC}
   \text{\footnotesize $\log\langle M_{GC}\rangle = 5.698 + 0.1294M_V^T + 0.0054(M_V^T)^2$}
\end{equation}
We adopt an uncertainty on our calculated $\log\langle M_{GC}\rangle$ values of $\pm 0.1$ dex based on the scatter in this relation around BCGs \citep{Harris14}. This mass range distinction is important as scatter increases for dwarf galaxies \citep{Villegas10}. The final resulting GCS masses for our sample can be found in Table \ref{tab:systems}.

\subsection{Determining Halo Masses}\label{sec:halo}

In this work we use the SHMR (stellar-to-halo mass ratio) as defined in \cite{Hudson15} to convert stellar masses to halo masses for the galaxies in our sample. This relation is well defined for massive galaxies such as the ones being studied here and is described by Equation \ref{eq:SHMR} below.
\begin{equation}\label{eq:SHMR}
     M_{\star}/M_h = 2 f_{1} \Bigg[\Big(\frac{M_{\star}}{M_1}\Big)^{-0.43}+\frac{M_{\star}}{M_1}\Bigg]^{-1} 
\end{equation}
Here $M_1$ is the transition or pivot halo mass, set to $10^{10.76} M_{\odot}$, and $f_{1}$ is the mass ratio at $M_1$, which is $f_1 = 0.0227$. This equation has been adjusted to a redshift of zero, which is appropriate given the low distances of the galaxies in this sample.

The stellar masses of the galaxies were calculated from their K-band total luminosities using the K-band stellar mass-to-light ratio, assuming a Chabrier/Kroupa mass function \citep{Bell03}.

\begin{center}
\begin{table*}[h!tb]
    \centering
    \caption{Final $M_{GCS}$ and $M_h$ Values} \label{tab:systems}
    \begin{tabular}{cccccc}
    \hline \hline
    Target Name & $N_{GC}$ & $R_{in}(kpc)$ & $R_{GCS}(kpc)$ & $M_{GCS}(\times 10^9 M_{\odot})$ & $M_h (\times 10^{13} M_{\odot})$ \\
    (1) & (2) & (3) & (4) & (5) & (6) \\
    \hline
    J13481399-3322547 & $1600 \pm 300$ & $5.6\pm0.2$ & $53.3\pm0.2$ & $0.42 \pm 0.08$ & $1.73\pm0.39$\\
    J13280261-3145207 & $3500 \pm 500$ & $5.5\pm0.2$ & $49.3\pm0.3$ & $1.03 \pm 0.17$ & $1.23 \pm 0.29$\\
    J13275493-3132187 & $10000 \pm 2000$ & $7.1\pm0.3$ & $78.1\pm0.4$ & $4.13 \pm 0.88$ & $4.89 \pm 1.14$\\
    J13272961-3123237 & $9100 \pm 1600$ & $11.6\pm0.5$ & $85.6\pm0.4$ & $3.76 \pm 0.74$ & $6.30 \pm 1.46$\\
    ESO 509-G067 & $4200 \pm 600$ & $4.8\pm0.2$ & $66.7\pm0.2$ & $1.73 \pm 0.28$ & $2.92 \pm 0.65$ \\
    ESO 509-G020 & $5800 \pm 800$ & $4.7\pm0.2$ & $68.3\pm0.2$ & $2.35 \pm 0.38 $ & $3.22 \pm 0.70$ \\
    ESO 509-G008 & $6500 \pm 900$ &  $5.6\pm0.2$ &$78.5\pm0.2$ & $2.45 \pm 0.38$ & $4.91 \pm 1.06$ \\
    ESO 444-G046 & $40800 \pm 6700$ & $14.9\pm0.6$ & $141\pm0.6$ & $26.4 \pm 4.80$ & $28.5 \pm 6.39$\\
    ESO 383-G076 & $39300 \pm 5000$ & $4.1\pm0.2$ & $124\pm0.4$ & $21.3 \pm 3.38$ & $18.6 \pm 3.95$\\
    ESO 325-G016 & $2800 \pm 400$ & $5.2\pm0.2$ & $54.1\pm0.2$ & $0.89 \pm 0.16$ & $1.53 \pm 0.34$ \\
    ESO 325-G004 & $9000\pm1100$ & $5.4\pm0.2$ & $85.4\pm0.2$ & $3.65 \pm 0.56$ & $6.20 \pm 1.34$\\
    ESO 306-G017 & $17000\pm6000$ & $10.5\pm0.5$ & $96.7\pm0.3$ & $9.24 \pm 3.36$ & $10.3 \pm 2.51$\\
    NGC 1129 & $7800\pm1500$ & $3.5\pm0.1$ & $73.8\pm0.5$ & $2.88 \pm 0.60$ & $4.59 \pm 0.97$\\
    NGC 1132 & $5600\pm1300$ & $2.8\pm0.1$ & $59.2\pm0.8$ & $1.85 \pm 0.46$ & $2.37 \pm 0.51$\\
    NGC 1272 & $6800\pm1500$ & $2.4\pm0.1$ & $57.7\pm0.8$ & $2.79 \pm 0.68$ & $2.19 \pm 0.47$\\
    NGC 1278 & $3300\pm900$ & $1.8\pm0.1$ & $44.7\pm1.3$ & $1.04 \pm 0.31$ & $1.02 \pm 0.21$\\
    NGC 3258 & $4800\pm600$ & $1.7\pm0.1$ & $42.4\pm1.5$ & $1.44 \pm 0.23$ & $0.87 \pm 0.17$\\
    NGC 3268 & $4300\pm600$ & $1.9\pm0.1$ & $46.1\pm1.2$ & $1.71 \pm 0.27$ & $1.12 \pm 0.23$\\
    NGC 3348 & $4000\pm500$ & $1.6\pm0.1$ & $42.8\pm1.4$ & $1.09 \pm 0.15$ & $0.89 \pm 0.18$\\
    NGC 4696 & $23300\pm3400$ & $2.8\pm0.1$ & $86.4\pm0.4$ & $12.5 \pm 2.11$ & $7.37 \pm 1.55$\\
    NGC 4874 & $12300\pm2800$ & $4.3\pm0.2$ & $81.4\pm0.4$ & $5.70 \pm 1.38$ & $6.15 \pm 1.31$\\
    NGC 4889 & $12600\pm3000$ & $5.0\pm0.2$ & $106.5\pm0.2$ & $5.93 \pm 1.51$ & $13.8 \pm 2.80$\\
    NGC 6166 & $18900\pm5400$ & $3.5\pm0.2$ & $91.8\pm0.3$ & $8.55 \pm 2.54$ & $8.85 \pm 1.88$\\
    NGC 7626 & $5100\pm600$ & $2.2\pm0.1$ & $53.7 \pm0.9$ & $1.64 \pm 0.23$ & $1.77 \pm 0.37$\\
    NGC 7720 & $15000\pm3700$ & $3.9\pm0.2$ & $80.9\pm0.4$ & $6.02 \pm 1.55$ & $6.04 \pm 1.29$\\
    UGC 9799 & $17000\pm4700$ & $3.5\pm0.1$ & $63.5\pm0.6$ & $7.68 \pm 2.22$ & $7.56 \pm 1.77$\\
    UGC 10143 & $8600\pm2500$ & $4.3\pm0.2$ & $87.2\pm0.3$ & $4.86 \pm 1.47$ & $2.94 \pm 0.65$\\
    \hline
    \end{tabular}
\item{} \footnotesize{\textit{Key to columns:} (1) Target name; (2) Total number of GCs; (3) Inner limit of GC photometry; (4) Standardized GCS radius; (4) GCS masses; (5) Dark matter halo masses.}
\end{table*}
\end{center}

\section{Results}\label{sec:results}

\subsection{Updated GCS Masses}\label{sec:masses}

\begin{figure}[h!tb]
    \includegraphics[width=0.45\textwidth]{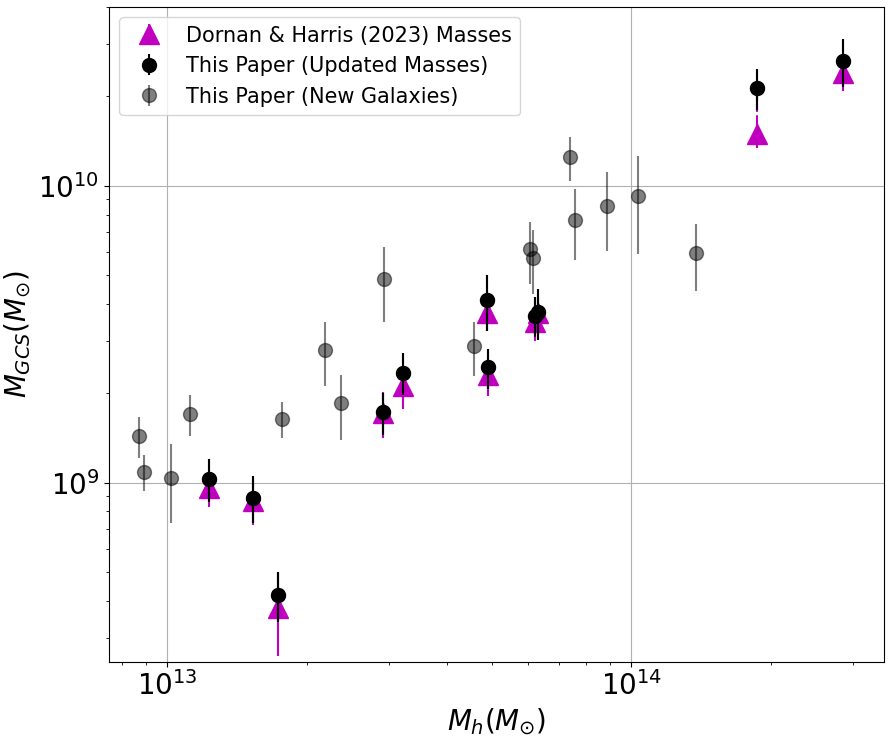}
    \caption{\label{fig:change} \footnotesize{$M_{GCS}-M_h$ relation plotted in log-log space for the galaxies studied in this paper. The $M_{GCS}$ values calculated using an annulus method and published in \cite{Dornan23} are plotted in purple triangles and the updated values calculated using the Voronoi method detailed int this paper are plotted in black circles. Grey circles denote the masses determined with the Voronoi method for the galaxies studied in this paper but not in \cite{Dornan23}.}}
\end{figure}

Figure \ref{fig:change} shows the GCS and halo masses for all the galaxies studied in this work. For the 11 galaxies which had masses previously determined in \cite{Dornan23} with the annulus method, we plot those old values in purple and the new masses in black.

As can be seen clearly in Figure \ref{fig:change} the updated masses are all of higher values and, with the exception of ESO 383-G076, within the errorbars of the original mass estimates. In addition, all updated GC radial density profile fits have lower uncertainties when calculated with the Voronoi method than with the annulus method, although due to the previously discussed higher ICGC uncertainty adopted in this paper, it does not end up translating to the $M_{GCS}$ uncertainties. 

The higher precision in the radial density profile fits with the Voronoi method was as expected based on the results of Paper I but, at first glance, the higher mass estimates were unexpected. Paper I found that for GCSs with more than $\sim 5000$ objects, as is the case for all of these massive elliptical galaxies, the Voronoi method should return an accurate $N_{GC}$ estimate and the annulus method should over-predict that value, not under-predict. However, upon further comparison of the simulated systems used in Paper I and the observed GCSs of the galaxies in this sample, it was found that although the sizes of these systems are similar, the steepnesses of their GC density profiles differ.

The massive simulated systems in Paper I all have shllower GCS profiles than 10 of the 11 GCSs studied with both methodologies in this paper. One simulated system in Paper I had a steeper profile than the galaxies in this paper, and it was found that the annulus method underpredicted its $N_{GC}$ estimate compared to the Voronoi method, which is what we see for these galaxies. This is due to the fact that for shallow profiles small areas of high local density in the inner regions of a distribution can artificially increase the average density of the innermost annuli, whereas for steep profiles the annulus bins cannot accurately mimic the more rapid change in radial density.

This steep simulated system in Paper I, however, was of a much smaller size than our observed galaxies, hosting only 300 objects. As such, the Voronoi method was found to predict the correct $N_{GC}$ only slightly better than the annulus method, likely due to small number statistics. We can extrapolate the results from Paper I out to massive systems with steep profiles; expecting the Voronoi method to accurately estimate $N_{GC}$ and the annulus method to underpredict it. 

ESO 383-G076 has an updated mass higher than calculated in \cite{Dornan23} despite also having a very shallow density profile with a powerlaw exponent of 0.615. This is because this galaxy is highly elliptical and this was not properly corrected for when determining the GC density profile in the previous paper. Thus, the increase in the mass estimate comes from taking ellipticity into account, rather than differences between the Voronoi and the annulus method.

\begin{figure*}[h!tb]
    \begin{center}
    \includegraphics[width=0.95\textwidth]{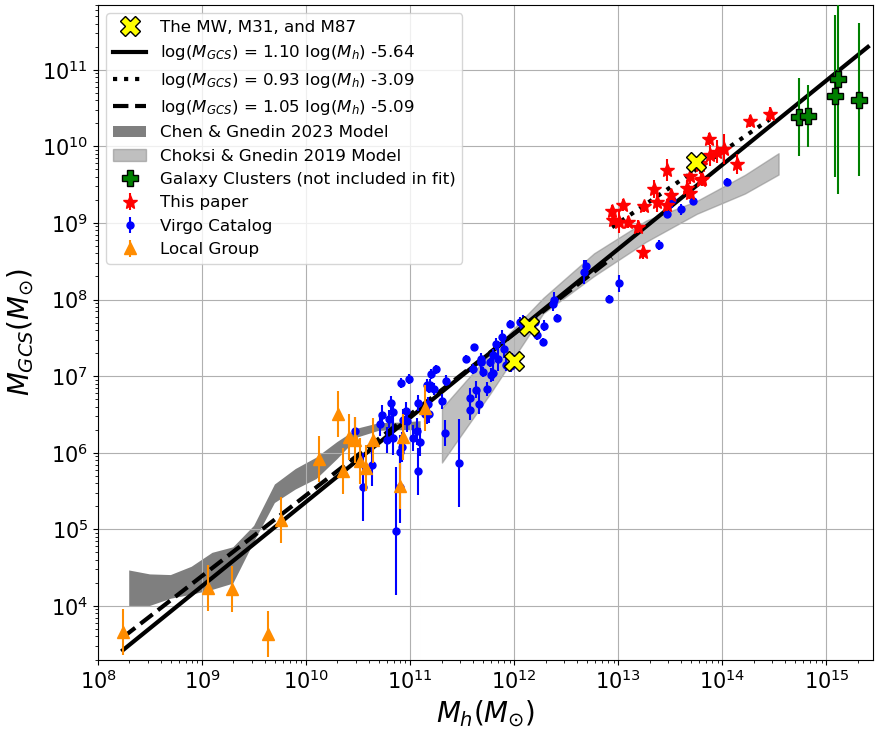}
    \caption{\label{fig:relation} \footnotesize{$M_{GCS}-M_h$ relation plotted in log-log space. Blue circles are Virgo catalog galaxies from \cite{Peng08}, orange triangles are low-mass local group galaxies, green crosses are galaxy clusters (not used for the fit), and red stars are the galaxies from this sample. The Milky Way, M31, and M87 have been highlighted in yellow crosses. The lighter shaded region is the results from the \cite{Choksi19} model and the darker shaded region is the results from the \cite{Chen23} model. The solid black line is the linear fit to all three extragalactic observational samples, the dashed line is the fit to just the local group and Virgo galaxy samples, and the dotted line in the fit to just the massive galaxy sample detailed in this paper.}}
    \end{center}
\end{figure*}

\subsection{The Global $M_{GCS}-M_h$ Relation} \label{sec:relation}

Once all of the masses for the galaxies in our sample were calculated they were plotted on the $M_{GCS}-M_h$ relation alongside galaxies of other mass ranges as well as predictions from simulations. This can be seen in Figure \ref{fig:relation}. For observational data we have plotted our sample of massive elliptical galaxies with galaxies in the ACS Virgo Cluster Survey \citep{Peng08} and the survey of low-mass galaxies in the Local Group which was used in \cite{Eadie22} and includes GCS data from \cite{Harris13, Forbes18, Forbes20}. We also highlight the locations of the Virgo BCG, M87 \citep{Peng08}, M31 \citep{Harris13, Patel17}, and the Milky Way \citep{Harris13, Kravtsov24} on the relation. 

We have also plotted where five galaxy clusters lie on the $M_{GCS}-M_h$ relation as well, represented by the green crosses in Figure \ref{fig:relation}, although these are not used in calculating the slope of the relation. These galaxy clusters include Abell 2744 \citep{Harris17}, Abell 1689 \citep{Alamo13,Harris17}, the Perseus cluster \citep{Harris17b}, the Coma cluster \citep{Peng11}, and the entirety of the Virgo cluster summed together \citep{Peng08, Durrell14}, and represent estimates of the $N_{GC}$ hosted both within and between all member galaxies, which results in the larger $M_{GCS}$ uncertainties. This estimate was done, very roughly, by using Equation \ref{eq:avgGC} to determine the $\langle M_{GC} \rangle$ for each cluster's BCG, assuming that approximately half of all the GCs hosted by the cluster have that mass, and that the other half have masses similar to the average Milky Way GC ($2.5\times 10^5 M_{\odot}$).

Rather than add in all galaxies with known GCSs (which make an inhomogeneous list with measurements from a wide variety of methods and raw data), here we deliberately subselect only the ones from the homogeneous, well defined Virgo survey, and the Local Group members whose GC numbers are well determined and allow us to extend the relation to the lowest possible galaxy masses. With this combination of observational samples the $M_{GCS}-M_h$ relation now spans over 7 orders of magnitude in GCS mass, the most complete and methodologically consistent global analysis of this relation to date. 

\begin{figure}[h!tb]
    \includegraphics[width=0.49\textwidth]{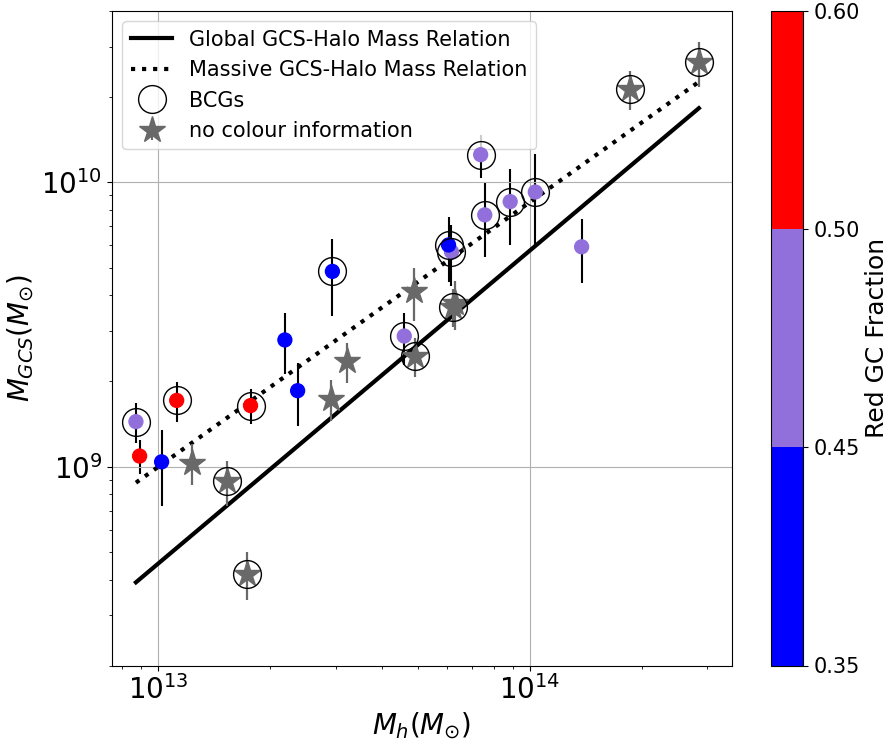}
    \includegraphics[width=0.49\textwidth]{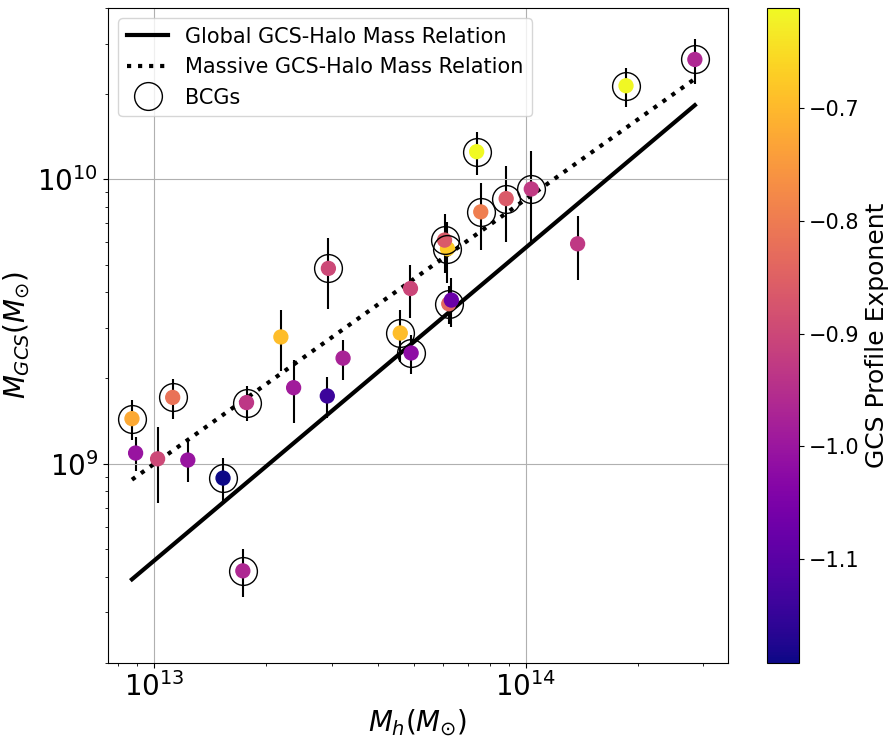}
    \caption{\label{fig:trends} \footnotesize{$M_{GCS}$ vs $M_h$ in log-log space for this paper's sample of massive galaxies. The linear fit determined from just this sample is represented by the dotted line, and the linear fit determined from this sample, the Virgo sample, and the Local Group sample is represented by the solid line. Black circles identify BCGs. \textit{Top:} Colour represents the fraction of red GCs, grey stars are galaxies without colour information. \textit{Bottom:} Colour represents the slope of the galaxies' GC radial density profile, with dark colours corresponding to steep and light corresponding to shallow.}}
\end{figure}

It should be noted that the uncertainties on the GCS masses for the Local Group dwarfs were adopted as 0.3 dex. Though the $N_{GC}$ values for the dwarfs are well known \citep{Eadie22}, the conversion to $M_{GCS}$ requires multiplying by a mean mass-to-light ratio, which observationally shows a typical cluster-to-cluster range of a factor of two \citep[cf.][]{Villegas10,Harris17}. Here we adopt a mean mass-to-light ratio of 1.4 for these local dwarfs. 

The shaded regions in Figure \ref{fig:relation} correspond to the results from \cite{Choksi19} and \cite{Chen23}. This data is results from analytical GC models applied to the \textit{Illustris-1-Dark} dark matter-only cosmological simulations. As can be seen in Figure \ref{fig:relation}, globally, the relation continues to hold linearly in log-log space across all three samples of galaxies. However, when fitting a linear regression to the observational data we found that the global relation is not only steeper than that predicted by the simulations, but is also slightly steeper than a 1:1 ratio. It should also be noted that when a linear regression is fit to only the sample of massive ellipticals studied here, of which the majority are BCGs, the slope is shallower than a 1:1 ratio. Table \ref{tab:fits} shows the solutions and associated uncertainties when fitting to each sample and combination of samples. Interestingly, we find that fitting only data from Local Group galaxies and M87 results in a close approximation of our fit to the full dataset. 

\begin{center}
\begin{table*}[h!tb]
    \centering
    \caption{Linear Fit Solutions for $M_{GCS}-M_h$ Relation} \label{tab:fits}
    \begin{tabular}{ccc}
    \hline \hline
    Sample Combination & Slope & Intercept \\
    (1) & (2) & (3) \\
    \hline
    All Samples & $1.10 \pm 0.02$ & $-5.64 \pm 0.89$\\
    \hline
    Massive Galaxies Only & $0.93\pm 0.09$ & $-3.09 \pm 1.02$\\
    Virgo Cluster Only & $1.05 \pm 0.03$ & $-5.05 \pm 0.91$\\
    Local Group Only & $1.11 \pm 0.15$ & $-5.82 \pm 1.04$\\
    \hline
    Massive Galaxies + Virgo Cluster & $1.10 \pm 0.03$ & $-5.62 \pm 0.89$\\
    Massive Galaxies + Local Group & $1.13 \pm 0.04$ & $-5.85 \pm 0.94$\\
    Virgo Cluster + Local Group & $1.05 \pm 0.03$ & $-5.09 \pm 0.90$\\
    \hline
    Local Group + M87 & $1.16 \pm  0.06$& $-6.32 \pm 0.96$ \\
    \hline
    \hline
    \end{tabular}
\item{} \footnotesize{\textit{Key to columns:} (1) Combination of observational samples used for the linear fits: either all of them, only one, or a combination of two; (2) The slope of the fit in log-log space; (3) The intercept of the fit in log-log space. }
\end{table*}
\end{center}

\subsection{Trends with GCS Properties}\label{sec:trends}

We also take the opportunity to study how two different GCS properties may relate to their offset from the main relation. Figure \ref{fig:trends} plots the massive galaxies in our sample with the colour bars representing the fraction of GCs that are red rather than blue (upper panel) and the exponent of the GC density profile as a representation of profile steepness (lower panel). We also plot the slope of the $M_{GCS}-M_h$ relation for all galaxies with a solid line, as was plotted in Figure \ref{fig:relation}, as well as the $M_{GCS}-M_h$ relation determined for just the massive galaxies in our sample with a dotted line. 

16 of the 27 galaxies in our sample have imaging with multiple filters available, allowing for color information. We find that the GCSs of these galaxies have bimodial colour distribution (see Figure 12-14 in \cite{Harris23}), which is expected for galaxies in this mass range \citep{Hartman23}. We define red vs blue GCs based on the double-Gaussian fit to the color distributions for each GCS, where the red vs blue fits become dominant over the other. We find no significant trend between galaxy position on the relation and GC fraction of red $N_{GC}$.

However, when we compare GC radial density profile steepness against position on the relation we \textit{do} see a trend. We find that, independent of host galaxy mass, a galaxy is more likely to lie above the $M_{GCS}-M_h$ relation if it has a shallow GC radial density profile and it is more likely to lie below the relation if it has a steep profile.

\begin{figure*}[h!tb]
    \begin{center}
    \includegraphics[width=0.99\textwidth]{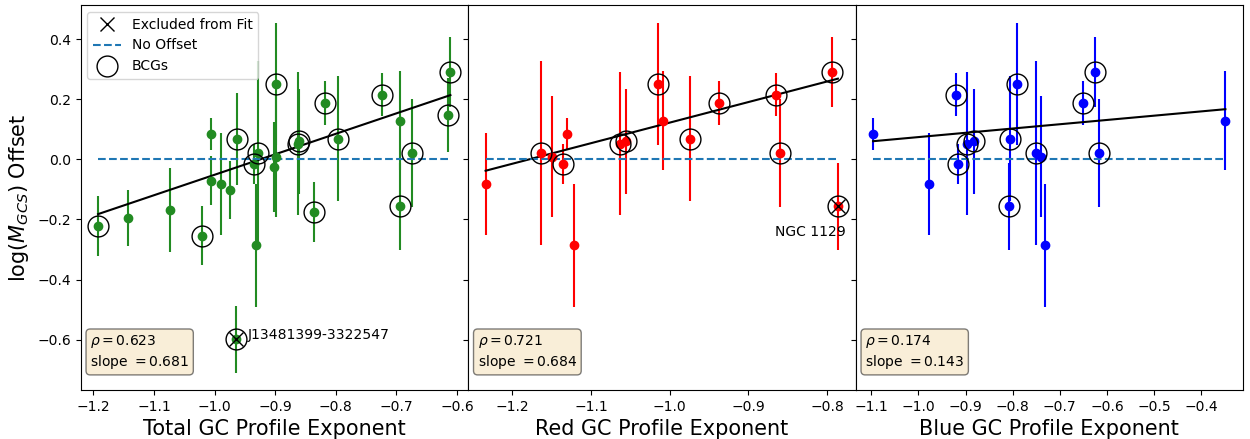}       \caption{\label{fig:offset} \footnotesize{All subfigures show the $\log(M_{GCS})$ offset from the $M_{GCS}-M_h$ relation derived from the massive galaxy sample as a function of slope of the galaxies' total GC density profiles, with more negative numbers being steeper and more positive numbers being more shallow. \textit{Left:} Total offset vs slope of density profile for all GCs. \textit{Middle:} Total offset vs slope of red GC density profile. \textit{Right:} Total offset vs slope of blue GC density profile. Note that the leftmost subfigure plots the full 27 galaxy sample, while the middle and right subfigures only plot the 16 galaxies with colour information.}}
    \end{center}
\end{figure*}

Figure \ref{fig:offset} shows the amount a galaxy is offset from the $M_{GCS}-M_h$ relation derived form the massive galaxy sample in log-space as a function of radial density profile exponent derived from all GCs in the system, red GCs only, and blue GCs only. The left panel plots all 27 massive ellipticals in our sample, while the middle and right panels plot only the 16 ellipticals with colour information. Figure \ref{fig:grid_colour} shows the fits for red and blue GCs separately for the galaxies in our sample with colour information, along with the profile exponents.

It can be clearly seen that this trend exists when looking at both the steepnesses derived from the total GCS and the red GCs, but is not present for the blue GCs. We found a Spearman correlation coefficient of 0.63, with a p-value of 0.0005, and a slope of 0.73 for total GCS profile steepness. When looking at the red and blue GC density profiles we find that this correlation is being driven entirely by the red GC population, as there exists no trend with relation offset and blue GC density profile steepness (Spearman correlation coefficient of 0.17, and a slope of 0.16). However, the correlation with relation offset and red GC density profile steepness is even tighter and steeper, with a Spearman correlation coefficient of 0.72, with a p-value of 0.002, and a slope of 0.68 when removing the outlier NGC 1129. 

We find that for the galaxies studied in this paper that whether a galaxy is classified as BCG or not does not have any significant affect on where it lies on the relation or on the trend with offset. This is likely because any non-BCGs studied in this sample are still NMCGs, with similar properties.

\section{Discussion}\label{sec:discussion}

\subsection{Comparison with Simulations}

First, let us compare the observational and theoretical $M_{GCS}-M_h$ relations shown in Figure \ref{fig:relation}. It can be seen that the \cite{Choksi19} model does an excellent job of predicting the positions of the Virgo Cluster galaxy members with halo masses greater than $10^{11} M_{\odot}$, with the exception of the BCG M87. However, its slope decreases past halo masses of $10^{13}M_{\odot}$ and the model ends up falling short of the GCS masses of the massive ellipticals (which are mostly BCGs and NMCGs) in our sample by about 0.3 dex. This offset is another representation of the long-known tendency of BCGs to have $2-3$ times higher specific frequencies compared with giant galaxies of the same luminosity but not located in central positions within their environments \citep[e.g.][]{Harris_vdb81,Peng08,Harris13}. The archetype of this difference is the pair of Virgo giants, M87 and M49, but another good example is the pairing of NGC 4874 and 4889 in Coma. Essentially, the \citet{Choksi19} model does well at matching the luminous non-BCGs in the list but underpredicts the BCGs.

A potential explanation for this discrepancy was offered by \cite{Choksi19}, as they note that it was found by \cite{Li19} that the GC initial mass function can shift to higher masses when the host galaxy experiences a major merger. This in turn increases the likelihood of much more massive GCs forming, which allows for more clusters to survive to present-day. However, the \cite{Choksi19} model used a merger-independent cluster formation rate, meaning they did not take this GC initial mass function shift into account. This would explain why the model would succeed in predicting the locations of lower-mass and non-BCG Virgo cluster members, as these galaxies have experienced comparatively less major mergers than the BCGs and NMCGs in our sample.

The \cite{Chen23} model, when compared to our Local Group observational sample here, appears to over-predict the GCS masses of these dwarf galaxies in Figure \ref{fig:relation}. However, this opposite effect appears to be at least in part because our Local Group sample, although chosen due to the higher confidence in the estimated masses, does not fully capture the extremely high scatter in the $M_{GCS}-M_h$ relation in this mass range. Figure 5 in \cite{Chen23} shows the model alongside data from \cite{Forbes18} which illustrates this clearly. The addition of many more dwarfs with carefully determined GC numbers would be an important step, and will be the subject of a later paper (see Section \ref{sec:future}).

\subsection{Drivers of Relation Offset}

 With the compilation of a homogeneous sample of massive elliptical galaxies in this paper, the extrinsic scatter in the $M_{GCS}-M_h$ relation has been minimized. Now we may investigate the intrinsic scatter in the relation to determine the physical processes pushing galaxies' GCS masses higher or lower than expected for their halo masses. The results from Figure \ref{fig:trends} shows that shallow density profiles for red GCs correlate the strongest with positive offset from the $M_{GCS}-M_h$ relation. 

 It has been seen observationally that GC metallicity, and by extent, colour, is related to age, across host galaxy scales \citep{Cote98, Katz14, Fahrion20, Escudero22}. In general, it is found that red GCs tend to be formed \textit{in-situ} and blue GCs \textit{ex-situ}, making their way into massive galaxies like the ones in this sample via accretion of smaller satellites. Because of this, it has also been found that the spatial and kinematic properties of both GC populations also differ, with blue GC profiles being more extended and shallow than the red GC profiles in the same galaxy \citep{Kluge23a, Belokurov24, Versic24}. As such, it is not surprising to find that the galaxies in our sample have differing red and blue GC density profiles slopes or trends with relation offset. 

 However, one would assume that since blue GCs tend to be formed \textit{ex-situ} they would trace rich merger histories more readily and thus be the population driving a trend with offset from the $M_{GCS}-M_h$ relation, not the red GCs. The result that we see instead implies that purely the number of mergers a galaxy has undergone does not determine its position on the relation, but rather the \textit{type} of mergers. The population of red GCs hosted by a massive elliptical can either be formed \textit{in-situ} or be accreted from a major merger with another massive elliptical which hosts many old, red GCs. Thus, the results of Figure \ref{fig:trends} implies that having experienced many minor mergers does little to drive a galaxy's GCS mass above what is expected for its halo mass, but a few major mergers will.

 This result is in agreement with what was found by \cite{Kluge23b}, who studied surface brightness profiles around BCGs, rather than GCS profiles. They found that the ``excess light" that BCGs host, making them more luminous than very massive normal elliptical galaxies, must be caused by major mergers with other massive ellipticals, with minor mergers playing a comparatively smaller role. 

 It should be noted that the galaxy NGC 1129 is excluded from the fit for this trend. This is the galaxy in our sample with the most shallow red GC density profile, yet it lies below the relation. NGC 1129 has been know to have a peculiar morphology for decades, with observations indicating a perpendicular light profile to its major axis \citep{Peletier90, Goullaud18, Ene20}, a twisted stellar kinematic profile to its major axis \citep{Veale17}, and a double-core at its centre \citep{Lyman16}. All of these observations indicate that NGC 1129 has undergone a recent, major merger, unlike the other galaxies in our sample.

 
 

 \subsection{Comparison of $M_{GCS}-M_h$ Slopes Between Galaxy Samples}

This interpretation of the cause of galaxy offset from the $M_{GCS}-M_h$ relation is supported by the differing slopes and intercepts of the relation for the Virgo cluster and massive galaxy samples. Table \ref{tab:fits} shows that when studied individually, both the Virgo cluster sample (comprised mostly of galaxies below the BCG mass regime) and the massive galaxy sample studied here have relatively similar slopes, close to a 1:1 ratio. However, their intercepts differ, with the massive galaxy sample $\log{(M_{GCS})}$ values sitting systematically higher than those for the Virgo cluster sample. When the linear fits of the two samples are compared using the Chow test \citep{chow_test}, we find a Chow statistic of 8.1 and a p-value of 0.0005, indicating that the difference in linear fits is statistically significant.

In this picture, our sample of massive galaxies have all experienced a higher number of mergers than the Virgo cluster galaxies, and of those mergers, the major mergers have accreted more massive ellitptical galaxies. Thus, systematically it would be expected that they would all have a positive offset from the global $M_{GCS}-M_h$ relation. In fact, it can be seen in Figure \ref{fig:relation} that M87, the Virgo cluster BCG, sits above all other Virgo galaxies, right alongside this paper's massive galaxy sample. Of the six non-BCG Virgo galaxies with  $M_h > 10^{13}M_{\odot}$ four seem to sit on the negative-most offset boundary of the massive galaxy sample (NGC 4649, NGC 4374, NGC 4365, and NGC 4406) and two appear to be more in line with the Virgo cluster sample relation (NGC 4472, NGC 4382). The four with similar positions to the massive galaxy sample have central positions in the Virgo A subcluster and could be considered NMCGs, or in the case of NGC 4365 is the central galaxy to a separate galaxy group \citep{Blom14}. On the other hand, NGC 4472 (M49), while also being a central galaxy and the Virgo sample galaxy with the highest $M_h$, sits at the centre of the smaller Virgo B subcluster. NGC 4382, the lowest mass Virgo galaxy with a $M_h$ still above $10^{13}M_{\odot}$, sits on the outskirts of the Virgo A subcluster, and would be considered a normal luminous elliptical rather than a BCG or NMCG.

It has been a subject of debate whether the high-mass end of the $M_{GCS}-M_h$ relation is genuinely linear or begins to display a downward curve \citep{Harris17, Boylan17,Choksi19,El-badry19}. The results of this paper would suggest that past work which have indicated a curvature of the $M_{GCS}-M_h$ relation may have simply used an observational sample of massive galaxies with fewer major mergers, or simulations which did not fully account for major mergers or did not use accurate GC disruption formation and disruption rates for BCGs. Both of these scenarios might result in only galaxies lying below $M_{GCS}-M_h$ relation and giving the appearance of a downward curve.

Interestingly, in Figure \ref{fig:relation} we can see that all five galaxy clusters are at the top end of the mass range but lie below the relation defined by the individual galaxies. This is expected, as the majority of GCs hosted by these massive clusters are either currently being hosted, or would have been originally formed in, smaller galaxies with lower specific GC frequencies \citep{Harris13, Moreno24}. Thus, these galaxy clusters' total GC specific frequencies are lower than for individual BCGs, putting them below the relation.

\subsection{Future Work}\label{sec:future}

In order to further investigate this difference in positions between the massive Virgo cluster sample galaxies and the massive galaxy sample studied in this paper, GCSs from very massive but \textit{non-central} galaxies must be studied. This would allow for a clearer understanding of how much merger history can drive galaxies above or below the relation.

In addition, a similar study to this one can be conducted with a sample of Milky Way mass galaxies to determine if there exists the same trend with GC profile slope and $M_{GCS}-M_h$ relation offset in lower mass regimes. For these lower-mass galaxies there will be other observational considerations to take into account due to their differing merger histories and mass contents from BCGs. For example, major mergers likely will not be depositing older, red populations of GCs, and thus any trend we observe would likely be with blue GCs instead. As well, GCs hosted by lower-mass galaxies also tend to be, on average, lower-mass themselves (see Equation \ref{eq:avgGC}) \citep{Villegas10,Harris13}. Thus, major mergers for Milky Way-like galaxies may disrupt more of the GC population than for massive ellipticals, but also due to their higher gas mass, at early enough times these mergers could also increase GC formation \citep{Newton24}.

Finally, it would also be greatly beneficial for a study such as this to be applied to a sample of dwarf galaxies. However, in order to determine the drivers of intrinsic scatter in this relation, extrinsic scatter from differing observational methods must first be minimized. The low-mass end of the $M_{GCS}-M_h$ relation is dominated by scatter \citep{Georgiev10, Forbes18, Berek24}, in part due to the observational difficulties in observing the complete GCSs of these small, dim galaxies, and in having enough luminous tracers at high enough radii to get accurate estimates of total halo masses. As well, there exists a diversity of dwarf galaxy morphologies, GC specific frequencies, and mass-to-light ratios that are likely also driving the intrinsic scatter in the relation \citep{Gannon22, Dokkum24, Li24}. The compilation of a literature catalog of dwarf galaxies with standardized GC counts and halo masses will be the subject of an upcoming paper. 

\section{Conclusions}\label{sec:conclusions}

In this work we studied the GCSs of 27 BCGs and NMCGs using HST data and a new Voronoi-tessellation based technique to determine accurate GC radial density profiles. For 16 of the 27 galaxies in our sample with colour information we determined the profiles for both the red and blue GC populations for each galaxy. We were able to plot this sample of massive galaxies alongside Virgo cluster galaxies and Local Group dwarf galaxies on the $M_{GCS}-M_h$ relation, spanning seven decades of GCS masses, the most complete observational view of this relation to date. From this, we found the following results:

\begin{enumerate}
    \item We found the $M_{GCS}-M_h$ relation across all galaxy masses to be slightly steeper than 1:1 linearity, with a slope of 1.10, however this is only present when connecting our BCG and non-BCG samples. The $M_{GCS}-M_h$ relation for non-BCG galaxies has a slope of 1.05.
    \item The nearly 1:1 linear $M_{GCS}-M_h$ relation holds for extremely massive, central galaxies, with a slope of 0.93. However, our massive galaxy sample is systematically shifted to higher GCS masses than for lower-mass galaxies. 
    \item There exists a negative trend with GCS radial density profile steepness and host galaxy offset from the $M_{GCS}-M_h$ relation, with galaxies with shallower GCS profiles being more likely to sit above the relation.
    \item This trend is even tighter and steeper when considering the profiles of only the red GCs hosted by galaxies, and no trend exists for the profiles of only the blue GCs hosted by galaxies.
    \item Thus, the red GC populations of the massive galaxies in our samples are driving this trend with offset, with the observed shallower red GC profiles having been caused by previous major mergers with other red-GC rich massive ellipticals.
    \item The lack of trend with offset for blue GCs suggests that minor mergers, which deposit more blue GCs, do little to affect massive ellipticals' positions on the $M_{GCS}-M_h$ relation.
    \item Therefore, we find that major mergers have the largest influence on intrinsic scatter around the $M_{GCS}-M_h$ relation for extremely massive, central, elliptical galaxies.
\end{enumerate}

\section{Acknowledgments}

We would like to acknowledge that this work made extensive use of the NASA/IPAC Extra-galactic Database and the SIMBAD database, operated at CDS, Strasbourg, France. We would like to thank the referee for their careful consideration and helpful comments on this paper.

This work is based on observations made with the NASA/ESA Hubble Space Telescope obtained from the Space Telescope Science Institute, which is operated by the Association of Universities for Research in Astronomy, Inc., under NASA contract NAS 5–26555. These observations can be accessed via doi: 10.17909/f01j-v360.

This work was supported by a Discovery Grant to WEH from the Natural Sciences and Engineering Research Council of Canada (NSERC). 

\bibliography{paper}{}

\begin{thebibliography}{}
\expandafter\ifx\csname natexlab\endcsname\relax\def\natexlab#1{#1}\fi
\providecommand{\url}[1]{\href{#1}{#1}}
\providecommand{\dodoi}[1]{doi:~\href{http://doi.org/#1}{\nolinkurl{#1}}}
\providecommand{\doeprint}[1]{\href{http://ascl.net/#1}{\nolinkurl{http://ascl.net/#1}}}
\providecommand{\doarXiv}[1]{\href{https://arxiv.org/abs/#1}{\nolinkurl{https://arxiv.org/abs/#1}}}

\bibitem[{{Alamo-Mart{\'\i}nez} {et~al.}(2013){Alamo-Mart{\'\i}nez}, {Blakeslee}, {Jee}, {C{\^o}t{\'e}}, {Ferrarese}, {Gonz{\'a}lez-L{\'o}pezlira}, {Jord{\'a}n}, {Meurer}, {Peng}, \& {West}}]{Alamo13}
{Alamo-Mart{\'\i}nez}, K.~A., {Blakeslee}, J.~P., {Jee}, M.~J., {et~al.} 2013, \apj, 775, 20, \dodoi{10.1088/0004-637X/775/1/20}

\bibitem[{{Beasley}(2020)}]{Beasley20}
{Beasley}, M.~A. 2020, in Reviews in Frontiers of Modern Astrophysics; From Space Debris to Cosmology, ed. P.~{Kab{\'a}th}, D.~{Jones}, \& M.~{Skarka}, 245--277, \dodoi{10.1007/978-3-030-38509-5_9}

\bibitem[{{Bell} {et~al.}(2003){Bell}, {McIntosh}, {Katz}, \& {Weinberg}}]{Bell03}
{Bell}, E.~F., {McIntosh}, D.~H., {Katz}, N., \& {Weinberg}, M.~D. 2003, Astrophysical Journal Supplement Series, 149, 289, \dodoi{10.1086/378847}

\bibitem[{{Belokurov} \& {Kravtsov}(2023)}]{Belokurov23}
{Belokurov}, V., \& {Kravtsov}, A. 2023, \mnras, 525, 4456, \dodoi{10.1093/mnras/stad2241}

\bibitem[{{Belokurov} \& {Kravtsov}(2024)}]{Belokurov24}
---. 2024, \mnras, 528, 3198, \dodoi{10.1093/mnras/stad3920}

\bibitem[{{Berek} {et~al.}(2024){Berek}, {Eadie}, {Speagle}, \& {Wang}}]{Berek24}
{Berek}, S.~C., {Eadie}, G.~M., {Speagle}, J.~S., \& {Wang}, S.~Y. 2024, \apj, 972, 104, \dodoi{10.3847/1538-4357/ad6147}

\bibitem[{{Blakeslee} {et~al.}(1997){Blakeslee}, {Tonry}, \& {Metzger}}]{Blakeslee97}
{Blakeslee}, J.~P., {Tonry}, J.~L., \& {Metzger}, M.~R. 1997, \aj, 114, 482, \dodoi{10.1086/118488}

\bibitem[{{Blom} {et~al.}(2014){Blom}, {Forbes}, {Foster}, {Romanowsky}, \& {Brodie}}]{Blom14}
{Blom}, C., {Forbes}, D.~A., {Foster}, C., {Romanowsky}, A.~J., \& {Brodie}, J.~P. 2014, \mnras, 439, 2420, \dodoi{10.1093/mnras/stu095}

\bibitem[{{Boylan-Kolchin}(2017)}]{Boylan17}
{Boylan-Kolchin}, M. 2017, \mnras, 472, 3120, \dodoi{10.1093/mnras/stx2164}

\bibitem[{{Burkert} \& {Forbes}(2020)}]{Burkert20}
{Burkert}, A., \& {Forbes}, D.~A. 2020, \aj, 159, 56, \dodoi{10.3847/1538-3881/ab5b0e}

\bibitem[{{Capuzzo-Dolcetta} \& {Mastrobuono-Battisti}(2009)}]{Capuzzo09}
{Capuzzo-Dolcetta}, R., \& {Mastrobuono-Battisti}, A. 2009, \aap, 507, 183, \dodoi{10.1051/0004-6361/200912255}

\bibitem[{{Chen} \& {Gnedin}(2023)}]{Chen23}
{Chen}, Y., \& {Gnedin}, O.~Y. 2023, \mnras, 522, 5638, \dodoi{10.1093/mnras/stad1328}

\bibitem[{{Choksi} \& {Gnedin}(2019)}]{Choksi19}
{Choksi}, N., \& {Gnedin}, O.~Y. 2019, \mnras, 488, 5409, \dodoi{10.1093/mnras/stz2097}

\bibitem[{Chow(1960)}]{chow_test}
Chow, G.~C. 1960, Econometrica, 28, 591

\bibitem[{{C{\^o}t{\'e}} {et~al.}(1998){C{\^o}t{\'e}}, {Marzke}, \& {West}}]{Cote98}
{C{\^o}t{\'e}}, P., {Marzke}, R.~O., \& {West}, M.~J. 1998, \apj, 501, 554, \dodoi{10.1086/305838}

\bibitem[{{Dolphin}(2000)}]{Dolphin00}
{Dolphin}, A.~E. 2000, \pasp, 112, 1383, \dodoi{10.1086/316630}

\bibitem[{{Dornan} \& {Harris}(2023)}]{Dornan23}
{Dornan}, V., \& {Harris}, W.~E. 2023, \apj, 950, 179, \dodoi{10.3847/1538-4357/accbc3}

\bibitem[{{Dornan} \& {Harris}(2024)}]{Dornan24}
---. 2024, \aj, 168, 48, \dodoi{10.3847/1538-3881/ad5447}

\bibitem[{{Durrell} {et~al.}(2014){Durrell}, {C{\^o}t{\'e}}, {Peng}, {Blakeslee}, {Ferrarese}, {Mihos}, {Puzia}, {Lan{\c{c}}on}, {Liu}, {Zhang}, {Cuillandre}, {McConnachie}, {Jord{\'a}n}, {Accetta}, {Boissier}, {Boselli}, {Courteau}, {Duc}, {Emsellem}, {Gwyn}, {Mei}, \& {Taylor}}]{Durrell14}
{Durrell}, P.~R., {C{\^o}t{\'e}}, P., {Peng}, E.~W., {et~al.} 2014, \apj, 794, 103, \dodoi{10.1088/0004-637X/794/2/103}

\bibitem[{{Eadie} {et~al.}(2022){Eadie}, {Harris}, \& {Springford}}]{Eadie22}
{Eadie}, G.~M., {Harris}, W.~E., \& {Springford}, A. 2022, \apj, 926, 162, \dodoi{10.3847/1538-4357/ac33b0}

\bibitem[{{El-Badry} {et~al.}(2019){El-Badry}, {Quataert}, {Weisz}, {Choksi}, \& {Boylan-Kolchin}}]{El-badry19}
{El-Badry}, K., {Quataert}, E., {Weisz}, D.~R., {Choksi}, N., \& {Boylan-Kolchin}, M. 2019, \mnras, 482, 4528, \dodoi{10.1093/mnras/sty3007}

\bibitem[{{Ene} {et~al.}(2020){Ene}, {Ma}, {Walsh}, {Greene}, {Thomas}, \& {Blakeslee}}]{Ene20}
{Ene}, I., {Ma}, C.-P., {Walsh}, J.~L., {et~al.} 2020, \apj, 891, 65, \dodoi{10.3847/1538-4357/ab7016}

\bibitem[{{Escudero} {et~al.}(2022){Escudero}, {Cortesi}, {Faifer}, {Sesto}, {Smith Castelli}, {Johnston}, {Reynaldi}, {Chies-Santos}, {Salinas}, {Men{\'e}ndez-Delmestre}, {Gon{\c{c}}alves}, {Grossi}, \& {Mendes de Oliveira}}]{Escudero22}
{Escudero}, C.~G., {Cortesi}, A., {Faifer}, F.~R., {et~al.} 2022, \mnras, 511, 393, \dodoi{10.1093/mnras/stac021}

\bibitem[{{Fahrion} {et~al.}(2020){Fahrion}, {Lyubenova}, {Hilker}, {van de Ven}, {Falc{\'o}n-Barroso}, {Leaman}, {Mart{\'\i}n-Navarro}, {Bittner}, {Coccato}, {Corsini}, {Gadotti}, {Iodice}, {McDermid}, {Pinna}, {Sarzi}, {Viaene}, {de Zeeuw}, \& {Zhu}}]{Fahrion20}
{Fahrion}, K., {Lyubenova}, M., {Hilker}, M., {et~al.} 2020, \aap, 637, A26, \dodoi{10.1051/0004-6361/202037685}

\bibitem[{{Federle} {et~al.}(2024){Federle}, {G{\'o}mez}, {Mieske}, {Harris}, {Hilker}, {Yegorova}, \& {Harris}}]{Federle24}
{Federle}, S., {G{\'o}mez}, M., {Mieske}, S., {et~al.} 2024, \aap, 689, A342, \dodoi{10.1051/0004-6361/202348235}

\bibitem[{{Forbes}(2017)}]{Forbes17}
{Forbes}, D.~A. 2017, \mnras, 472, L104, \dodoi{10.1093/mnrasl/slx148}

\bibitem[{{Forbes}(2020)}]{Forbes20}
---. 2020, \mnras, 493, 847, \dodoi{10.1093/mnras/staa245}

\bibitem[{{Forbes} {et~al.}(2018){Forbes}, {Read}, {Gieles}, \& {Collins}}]{Forbes18}
{Forbes}, D.~A., {Read}, J.~I., {Gieles}, M., \& {Collins}, M. L.~M. 2018, \mnras, 481, 5592, \dodoi{10.1093/mnras/sty2584}

\bibitem[{{Gannon} {et~al.}(2022){Gannon}, {Forbes}, {Romanowsky}, {Ferr{\'e}-Mateu}, {Couch}, {Brodie}, {Huang}, {Janssens}, \& {Okabe}}]{Gannon22}
{Gannon}, J.~S., {Forbes}, D.~A., {Romanowsky}, A.~J., {et~al.} 2022, \mnras, 510, 946, \dodoi{10.1093/mnras/stab3297}

\bibitem[{{Georgiev} {et~al.}(2010){Georgiev}, {Puzia}, {Goudfrooij}, \& {Hilker}}]{Georgiev10}
{Georgiev}, I.~Y., {Puzia}, T.~H., {Goudfrooij}, P., \& {Hilker}, M. 2010, \mnras, 406, 1967, \dodoi{10.1111/j.1365-2966.2010.16802.x}

\bibitem[{{Goullaud} {et~al.}(2018){Goullaud}, {Jensen}, {Blakeslee}, {Ma}, {Greene}, \& {Thomas}}]{Goullaud18}
{Goullaud}, C.~F., {Jensen}, J.~B., {Blakeslee}, J.~P., {et~al.} 2018, \apj, 856, 11, \dodoi{10.3847/1538-4357/aab1f3}

\bibitem[{{Harris}(2023)}]{Harris23}
{Harris}, W.~E. 2023, \apjs, 265, 9, \dodoi{10.3847/1538-4365/acab5c}

\bibitem[{{Harris} {et~al.}(2017){Harris}, {Blakeslee}, \& {Harris}}]{Harris17}
{Harris}, W.~E., {Blakeslee}, J.~P., \& {Harris}, G. L.~H. 2017, Astrophysical Journal, 836, 67, \dodoi{10.3847/1538-4357/836/1/67}

\bibitem[{{Harris} {et~al.}(2015){Harris}, {Harris}, \& {Hudson}}]{Harris15}
{Harris}, W.~E., {Harris}, G.~L., \& {Hudson}, M.~J. 2015, \apj, 806, 36, \dodoi{10.1088/0004-637X/806/1/36}

\bibitem[{{Harris} {et~al.}(2013){Harris}, {Harris}, \& {Alessi}}]{Harris13}
{Harris}, W.~E., {Harris}, G. L.~H., \& {Alessi}, M. 2013, \apj, 772, 82, \dodoi{10.1088/0004-637X/772/2/82}

\bibitem[{{Harris} \& {Mulholland}(2017)}]{Harris17b}
{Harris}, W.~E., \& {Mulholland}, C.~J. 2017, \apj, 839, 102, \dodoi{10.3847/1538-4357/aa6a59}

\bibitem[{{Harris} \& {van den Bergh}(1981)}]{Harris_vdb81}
{Harris}, W.~E., \& {van den Bergh}, S. 1981, \aj, 86, 1627, \dodoi{10.1086/113047}

\bibitem[{{Harris} {et~al.}(2014){Harris}, {Morningstar}, {Gnedin}, {O'Halloran}, {Blakeslee}, {Whitmore}, {C{\^o}t{\'e}}, {Geisler}, {Peng}, {Bailin}, {Rothberg}, {Cockcroft}, \& {Barber DeGraaff}}]{Harris14}
{Harris}, W.~E., {Morningstar}, W., {Gnedin}, O.~Y., {et~al.} 2014, Astrophysical Journal, 797, 128, \dodoi{10.1088/0004-637X/797/2/128}

\bibitem[{{Harris} {et~al.}(2020){Harris}, {Brown}, {Durrell}, {Romanowsky}, {Blakeslee}, {Brodie}, {Janssens}, {Lisker}, {Okamoto}, \& {Wittmann}}]{Harris20}
{Harris}, W.~E., {Brown}, R.~A., {Durrell}, P.~R., {et~al.} 2020, \apj, 890, 105, \dodoi{10.3847/1538-4357/ab6992}

\bibitem[{{Hartman} {et~al.}(2023){Hartman}, {Harris}, {Blakeslee}, {Ma}, \& {Greene}}]{Hartman23}
{Hartman}, K., {Harris}, W.~E., {Blakeslee}, J.~P., {Ma}, C.-P., \& {Greene}, J.~E. 2023, \apj, 953, 154, \dodoi{10.3847/1538-4357/ace340}

\bibitem[{{Hinshaw} {et~al.}(2013){Hinshaw}, {Larson}, {Komatsu}, {Spergel}, {Bennett}, {Dunkley}, {Nolta}, {Halpern}, {Hill}, {Odegard}, {Page}, {Smith}, {Weiland}, {Gold}, {Jarosik}, {Kogut}, {Limon}, {Meyer}, {Tucker}, {Wollack}, \& {Wright}}]{wmap13}
{Hinshaw}, G., {Larson}, D., {Komatsu}, E., {et~al.} 2013, \apjs, 208, 19, \dodoi{10.1088/0067-0049/208/2/19}

\bibitem[{{Hudson} {et~al.}(2014){Hudson}, {Harris}, \& {Harris}}]{Hudson14}
{Hudson}, M.~J., {Harris}, G.~L., \& {Harris}, W.~E. 2014, \apjl, 787, L5, \dodoi{10.1088/2041-8205/787/1/L5}

\bibitem[{{Hudson} \& {Robison}(2018)}]{Hudson18}
{Hudson}, M.~J., \& {Robison}, B. 2018, \mnras, 477, 3869, \dodoi{10.1093/mnras/sty844}

\bibitem[{{Hudson} {et~al.}(2015){Hudson}, {Gillis}, {Coupon}, {Hildebrandt}, {Erben}, {Heymans}, {Hoekstra}, {Kitching}, {Mellier}, {Miller}, {Van Waerbeke}, {Bonnett}, {Fu}, {Kuijken}, {Rowe}, {Schrabback}, {Semboloni}, {van Uitert}, \& {Velander}}]{Hudson15}
{Hudson}, M.~J., {Gillis}, B.~R., {Coupon}, J., {et~al.} 2015, Monthly Notices of the Royal Astronomical Society, 447, 298, \dodoi{10.1093/mnras/stu2367}

\bibitem[{{Huxor} {et~al.}(2011){Huxor}, {Ferguson}, {Tanvir}, {Irwin}, {Mackey}, {Ibata}, {Bridges}, {Chapman}, \& {Lewis}}]{Huxor11}
{Huxor}, A.~P., {Ferguson}, A.~M.~N., {Tanvir}, N.~R., {et~al.} 2011, \mnras, 414, 770, \dodoi{10.1111/j.1365-2966.2011.18450.x}

\bibitem[{{Joschko} {et~al.}(2024){Joschko}, {Kruijssen}, {Trujillo-Gomez}, {Pfeffer}, {Bastian}, {Crain}, \& {Reina-Campos}}]{Joschko24}
{Joschko}, P.~S., {Kruijssen}, J.~M.~D., {Trujillo-Gomez}, S., {et~al.} 2024, arXiv e-prints, arXiv:2412.04105, \dodoi{10.48550/arXiv.2412.04105}

\bibitem[{{Katz} \& {Ricotti}(2014)}]{Katz14}
{Katz}, H., \& {Ricotti}, M. 2014, \mnras, 444, 2377, \dodoi{10.1093/mnras/stu1489}

\bibitem[{{Kluge} \& {Bender}(2023)}]{Kluge23b}
{Kluge}, M., \& {Bender}, R. 2023, \apjs, 267, 41, \dodoi{10.3847/1538-4365/ace052}

\bibitem[{{Kluge} {et~al.}(2023){Kluge}, {Remus}, {Babyk}, {Forbes}, \& {Dolfi}}]{Kluge23a}
{Kluge}, M., {Remus}, R.-S., {Babyk}, I.~V., {Forbes}, D.~A., \& {Dolfi}, A. 2023, \mnras, 521, 4852, \dodoi{10.1093/mnras/stad882}

\bibitem[{{Kravtsov} \& {Winney}(2024)}]{Kravtsov24}
{Kravtsov}, A., \& {Winney}, S. 2024, The Open Journal of Astrophysics, 7, 50, \dodoi{10.33232/001c.120316}

\bibitem[{{Li} {et~al.}(2024){Li}, {Eadie}, {Brown}, {Harris}, {Abraham}, {van Dokkum}, {Janssens}, {Berek}, {Danieli}, {Romanowsky}, \& {Speagle}}]{Li24}
{Li}, D., {Eadie}, G., {Brown}, P., {et~al.} 2024, arXiv e-prints, arXiv:2409.06040, \dodoi{10.48550/arXiv.2409.06040}

\bibitem[{{Li} \& {Gnedin}(2019)}]{Li19}
{Li}, H., \& {Gnedin}, O.~Y. 2019, \mnras, 486, 4030, \dodoi{10.1093/mnras/stz1114}

\bibitem[{{Lim} {et~al.}(2024){Lim}, {Peng}, {C{\^o}t{\'e}}, {Ferrarese}, {Roediger}, {Liu}, {Spengler}, {Sola}, {Duc}, {Sales}, {Blakeslee}, {Cuillandre}, {Durrell}, {Emsellem}, {Gwyn}, {Lan{\c{c}}on}, {Marleau}, {Mihos}, {M{\"u}ller}, {Puzia}, \& {S{\'a}nchez-Janssen}}]{Lim24}
{Lim}, S., {Peng}, E.~W., {C{\^o}t{\'e}}, P., {et~al.} 2024, \apj, 966, 168, \dodoi{10.3847/1538-4357/ad3444}

\bibitem[{{Lyman} {et~al.}(2016){Lyman}, {Levan}, {James}, {Angus}, {Church}, {Davies}, \& {Tanvir}}]{Lyman16}
{Lyman}, J.~D., {Levan}, A.~J., {James}, P.~A., {et~al.} 2016, \mnras, 458, 1768, \dodoi{10.1093/mnras/stw477}

\bibitem[{{Madrid} {et~al.}(2018){Madrid}, {O'Neill}, {Gagliano}, \& {Marvil}}]{Madrid18}
{Madrid}, J.~P., {O'Neill}, C.~R., {Gagliano}, A.~T., \& {Marvil}, J.~R. 2018, \apj, 867, 144, \dodoi{10.3847/1538-4357/aae206}

\bibitem[{{Mirabile} {et~al.}(2024){Mirabile}, {Cantiello}, {Lonare}, {Ragusa}, {Paolillo}, {Hazra}, {La Marca}, {Iodice}, {Spavone}, {Mieske}, {Rejkuba}, {Hilker}, {Riccio}, {Habas}, {Brocato}, {Schipani}, {Grado}, \& {Limatola}}]{Mirabile24}
{Mirabile}, M., {Cantiello}, M., {Lonare}, P., {et~al.} 2024, \aap, 691, A104, \dodoi{10.1051/0004-6361/202451273}

\bibitem[{{Moreno-Hilario} {et~al.}(2024){Moreno-Hilario}, {Martinez-Medina}, {Li}, {Souza}, \& {P{\'e}rez-Villegas}}]{Moreno24}
{Moreno-Hilario}, E., {Martinez-Medina}, L.~A., {Li}, H., {Souza}, S.~O., \& {P{\'e}rez-Villegas}, A. 2024, \mnras, 527, 2765, \dodoi{10.1093/mnras/stad3306}

\bibitem[{{Newton} {et~al.}(2024){Newton}, {Davies}, {Pfeffer}, {Crain}, {Kruijssen}, {Pontzen}, \& {Bastian}}]{Newton24}
{Newton}, O., {Davies}, J.~J., {Pfeffer}, J., {et~al.} 2024, arXiv e-prints, arXiv:2409.04516, \dodoi{10.48550/arXiv.2409.04516}

\bibitem[{{Patel} {et~al.}(2017){Patel}, {Besla}, \& {Mandel}}]{Patel17}
{Patel}, E., {Besla}, G., \& {Mandel}, K. 2017, \mnras, 468, 3428, \dodoi{10.1093/mnras/stx698}

\bibitem[{{Peletier} {et~al.}(1990){Peletier}, {Davies}, {Illingworth}, {Davis}, \& {Cawson}}]{Peletier90}
{Peletier}, R.~F., {Davies}, R.~L., {Illingworth}, G.~D., {Davis}, L.~E., \& {Cawson}, M. 1990, \aj, 100, 1091, \dodoi{10.1086/115582}

\bibitem[{{Peng} {et~al.}(2008){Peng}, {Jord{\'a}n}, {C{\^o}t{\'e}}, {Takamiya}, {West}, {Blakeslee}, {Chen}, {Ferrarese}, {Mei}, {Tonry}, \& {West}}]{Peng08}
{Peng}, E.~W., {Jord{\'a}n}, A., {C{\^o}t{\'e}}, P., {et~al.} 2008, \apj, 681, 197, \dodoi{10.1086/587951}

\bibitem[{{Peng} {et~al.}(2011){Peng}, {Ferguson}, {Goudfrooij}, {Hammer}, {Lucey}, {Marzke}, {Puzia}, {Carter}, {Balcells}, {Bridges}, {Chiboucas}, {del Burgo}, {Graham}, {Guzm{\'a}n}, {Hudson}, {Matkovi{\'c}}, {Merritt}, {Miller}, {Mouhcine}, {Phillipps}, {Sharples}, {Smith}, {Tully}, \& {Verdoes Kleijn}}]{Peng11}
{Peng}, E.~W., {Ferguson}, H.~C., {Goudfrooij}, P., {et~al.} 2011, \apj, 730, 23, \dodoi{10.1088/0004-637X/730/1/23}

\bibitem[{{Planck Collaboration} {et~al.}(2016){Planck Collaboration}, {Ade}, {Aghanim}, {Arnaud}, {Ashdown}, {Aumont}, {Baccigalupi}, {Banday}, {Barreiro}, {Bartlett}, {Bartolo}, {Battaner}, {Battye}, {Benabed}, {Beno{\^\i}t}, {Benoit-L{\'e}vy}, {Bernard}, {Bersanelli}, {Bielewicz}, {Bock}, {Bonaldi}, {Bonavera}, {Bond}, {Borrill}, {Bouchet}, {Boulanger}, {Bucher}, {Burigana}, {Butler}, {Calabrese}, {Cardoso}, {Catalano}, {Challinor}, {Chamballu}, {Chary}, {Chiang}, {Chluba}, {Christensen}, {Church}, {Clements}, {Colombi}, {Colombo}, {Combet}, {Coulais}, {Crill}, {Curto}, {Cuttaia}, {Danese}, {Davies}, {Davis}, {de Bernardis}, {de Rosa}, {de Zotti}, {Delabrouille}, {D{\'e}sert}, {Di Valentino}, {Dickinson}, {Diego}, {Dolag}, {Dole}, {Donzelli}, {Dor{\'e}}, {Douspis}, {Ducout}, {Dunkley}, {Dupac}, {Efstathiou}, {Elsner}, {En{\ss}lin}, {Eriksen}, {Farhang}, {Fergusson}, {Finelli}, {Forni}, {Frailis}, {Fraisse}, {Franceschi}, {Frejsel}, {Galeotta}, {Galli}, {Ganga}, {Gauthier}, {Gerbino}, {Ghosh}, {Giard},
  {Giraud-H{\'e}raud}, {Giusarma}, {Gjerl{\o}w}, {Gonz{\'a}lez-Nuevo}, {G{\'o}rski}, {Gratton}, {Gregorio}, {Gruppuso}, {Gudmundsson}, {Hamann}, {Hansen}, {Hanson}, {Harrison}, {Helou}, {Henrot-Versill{\'e}}, {Hern{\'a}ndez-Monteagudo}, {Herranz}, {Hildebrandt}, {Hivon}, {Hobson}, {Holmes}, {Hornstrup}, {Hovest}, {Huang}, {Huffenberger}, {Hurier}, {Jaffe}, {Jaffe}, {Jones}, {Juvela}, {Keih{\"a}nen}, {Keskitalo}, {Kisner}, {Kneissl}, {Knoche}, {Knox}, {Kunz}, {Kurki-Suonio}, {Lagache}, {L{\"a}hteenm{\"a}ki}, {Lamarre}, {Lasenby}, {Lattanzi}, {Lawrence}, {Leahy}, {Leonardi}, {Lesgourgues}, {Levrier}, {Lewis}, {Liguori}, {Lilje}, {Linden-V{\o}rnle}, {L{\'o}pez-Caniego}, {Lubin}, {Mac{\'\i}as-P{\'e}rez}, {Maggio}, {Maino}, {Mandolesi}, {Mangilli}, {Marchini}, {Maris}, {Martin}, {Martinelli}, {Mart{\'\i}nez-Gonz{\'a}lez}, {Masi}, {Matarrese}, {McGehee}, {Meinhold}, {Melchiorri}, {Melin}, {Mendes}, {Mennella}, {Migliaccio}, {Millea}, {Mitra}, {Miville-Desch{\^e}nes}, {Moneti}, {Montier}, {Morgante}, {Mortlock},
  {Moss}, {Munshi}, {Murphy}, {Naselsky}, {Nati}, {Natoli}, {Netterfield}, {N{\o}rgaard-Nielsen}, {Noviello}, {Novikov}, {Novikov}, {Oxborrow}, {Paci}, {Pagano}, {Pajot}, {Paladini}, {Paoletti}, {Partridge}, {Pasian}, {Patanchon}, {Pearson}, {Perdereau}, {Perotto}, {Perrotta}, {Pettorino}, {Piacentini}, {Piat}, {Pierpaoli}, {Pietrobon}, {Plaszczynski}, {Pointecouteau}, {Polenta}, {Popa}, {Pratt}, \& {Pr{\'e}zeau}}]{planck15}
{Planck Collaboration}, {Ade}, P.~A.~R., {Aghanim}, N., {et~al.} 2016, \aap, 594, A13, \dodoi{10.1051/0004-6361/201525830}

\bibitem[{{Planck Collaboration} {et~al.}(2020){Planck Collaboration}, {Aghanim}, {Akrami}, {Ashdown}, {Aumont}, {Baccigalupi}, {Ballardini}, {Banday}, {Barreiro}, {Bartolo}, {Basak}, {Battye}, {Benabed}, {Bernard}, {Bersanelli}, {Bielewicz}, {Bock}, {Bond}, {Borrill}, {Bouchet}, {Boulanger}, {Bucher}, {Burigana}, {Butler}, {Calabrese}, {Cardoso}, {Carron}, {Challinor}, {Chiang}, {Chluba}, {Colombo}, {Combet}, {Contreras}, {Crill}, {Cuttaia}, {de Bernardis}, {de Zotti}, {Delabrouille}, {Delouis}, {Di Valentino}, {Diego}, {Dor{\'e}}, {Douspis}, {Ducout}, {Dupac}, {Dusini}, {Efstathiou}, {Elsner}, {En{\ss}lin}, {Eriksen}, {Fantaye}, {Farhang}, {Fergusson}, {Fernandez-Cobos}, {Finelli}, {Forastieri}, {Frailis}, {Fraisse}, {Franceschi}, {Frolov}, {Galeotta}, {Galli}, {Ganga}, {G{\'e}nova-Santos}, {Gerbino}, {Ghosh}, {Gonz{\'a}lez-Nuevo}, {G{\'o}rski}, {Gratton}, {Gruppuso}, {Gudmundsson}, {Hamann}, {Handley}, {Hansen}, {Herranz}, {Hildebrandt}, {Hivon}, {Huang}, {Jaffe}, {Jones}, {Karakci}, {Keih{\"a}nen},
  {Keskitalo}, {Kiiveri}, {Kim}, {Kisner}, {Knox}, {Krachmalnicoff}, {Kunz}, {Kurki-Suonio}, {Lagache}, {Lamarre}, {Lasenby}, {Lattanzi}, {Lawrence}, {Le Jeune}, {Lemos}, {Lesgourgues}, {Levrier}, {Lewis}, {Liguori}, {Lilje}, {Lilley}, {Lindholm}, {L{\'o}pez-Caniego}, {Lubin}, {Ma}, {Mac{\'\i}as-P{\'e}rez}, {Maggio}, {Maino}, {Mandolesi}, {Mangilli}, {Marcos-Caballero}, {Maris}, {Martin}, {Martinelli}, {Mart{\'\i}nez-Gonz{\'a}lez}, {Matarrese}, {Mauri}, {McEwen}, {Meinhold}, {Melchiorri}, {Mennella}, {Migliaccio}, {Millea}, {Mitra}, {Miville-Desch{\^e}nes}, {Molinari}, {Montier}, {Morgante}, {Moss}, {Natoli}, {N{\o}rgaard-Nielsen}, {Pagano}, {Paoletti}, {Partridge}, {Patanchon}, {Peiris}, {Perrotta}, {Pettorino}, {Piacentini}, {Polastri}, {Polenta}, {Puget}, {Rachen}, {Reinecke}, {Remazeilles}, {Renzi}, {Rocha}, {Rosset}, {Roudier}, {Rubi{\~n}o-Mart{\'\i}n}, {Ruiz-Granados}, {Salvati}, {Sandri}, {Savelainen}, {Scott}, {Shellard}, {Sirignano}, {Sirri}, {Spencer}, {Sunyaev}, {Suur-Uski}, {Tauber}, {Tavagnacco},
  {Tenti}, {Toffolatti}, {Tomasi}, {Trombetti}, {Valenziano}, {Valiviita}, {Van Tent}, {Vibert}, {Vielva}, {Villa}, {Vittorio}, {Wandelt}, {Wehus}, {White}, {White}, {Zacchei}, \& {Zonca}}]{planck18}
{Planck Collaboration}, {Aghanim}, N., {Akrami}, Y., {et~al.} 2020, \aap, 641, A6, \dodoi{10.1051/0004-6361/201833910}

\bibitem[{{Reina-Campos} {et~al.}(2023){Reina-Campos}, {Trujillo-Gomez}, {Pfeffer}, {Sills}, {Deason}, {Crain}, \& {Kruijssen}}]{Marta23}
{Reina-Campos}, M., {Trujillo-Gomez}, S., {Pfeffer}, J.~L., {et~al.} 2023, \mnras, 521, 6368, \dodoi{10.1093/mnras/stad920}

\bibitem[{{S{\'a}nchez-Janssen} {et~al.}(2019){S{\'a}nchez-Janssen}, {C{\^o}t{\'e}}, {Ferrarese}, {Peng}, {Roediger}, {Blakeslee}, {Emsellem}, {Puzia}, {Spengler}, {Taylor}, {{\'A}lamo-Mart{\'\i}nez}, {Boselli}, {Cantiello}, {Cuillandre}, {Duc}, {Durrell}, {Gwyn}, {MacArthur}, {Lan{\c{c}}on}, {Lim}, {Liu}, {Mei}, {Miller}, {Mu{\~n}oz}, {Mihos}, {Paudel}, {Powalka}, \& {Toloba}}]{NGVS20}
{S{\'a}nchez-Janssen}, R., {C{\^o}t{\'e}}, P., {Ferrarese}, L., {et~al.} 2019, \apj, 878, 18, \dodoi{10.3847/1538-4357/aaf4fd}

\bibitem[{{Spitler} \& {Forbes}(2009)}]{Spitler09}
{Spitler}, L.~R., \& {Forbes}, D.~A. 2009, \mnras, 392, L1, \dodoi{10.1111/j.1745-3933.2008.00567.x}

\bibitem[{{Valenzuela} {et~al.}(2024){Valenzuela}, {Remus}, {McKenzie}, \& {Forbes}}]{Valenzuela24}
{Valenzuela}, L.~M., {Remus}, R.-S., {McKenzie}, M., \& {Forbes}, D.~A. 2024, \aap, 687, A104, \dodoi{10.1051/0004-6361/202348010}

\bibitem[{{van Dokkum} {et~al.}(2024){van Dokkum}, {Li}, {Abraham}, {Danieli}, {Eadie}, {Harris}, \& {Romanowsky}}]{Dokkum24}
{van Dokkum}, P., {Li}, D.~D., {Abraham}, R., {et~al.} 2024, Research Notes of the American Astronomical Society, 8, 135, \dodoi{10.3847/2515-5172/ad4be6}

\bibitem[{{VandenBerg} {et~al.}(2013){VandenBerg}, {Brogaard}, {Leaman}, \& {Casagrande}}]{Vandenberg13}
{VandenBerg}, D.~A., {Brogaard}, K., {Leaman}, R., \& {Casagrande}, L. 2013, \apj, 775, 134, \dodoi{10.1088/0004-637X/775/2/134}

\bibitem[{{Veale} {et~al.}(2017){Veale}, {Ma}, {Thomas}, {Greene}, {McConnell}, {Walsh}, {Ito}, {Blakeslee}, \& {Janish}}]{Veale17}
{Veale}, M., {Ma}, C.-P., {Thomas}, J., {et~al.} 2017, \mnras, 464, 356, \dodoi{10.1093/mnras/stw2330}

\bibitem[{{Ver{\v{s}}i{\v{c}}} {et~al.}(2024){Ver{\v{s}}i{\v{c}}}, {Rejkuba}, {Arnaboldi}, {Gerhard}, {Pulsoni}, {Valenzuela}, {Hartke}, {Watkins}, {van de Ven}, \& {Thater}}]{Versic24}
{Ver{\v{s}}i{\v{c}}}, T., {Rejkuba}, M., {Arnaboldi}, M., {et~al.} 2024, \aap, 687, A80, \dodoi{10.1051/0004-6361/202349097}

\bibitem[{{Villegas} {et~al.}(2010){Villegas}, {Jord{\'a}n}, {Peng}, {Blakeslee}, {C{\^o}t{\'e}}, {Ferrarese}, {Kissler-Patig}, {Mei}, {Infante}, {Tonry}, \& {West}}]{Villegas10}
{Villegas}, D., {Jord{\'a}n}, A., {Peng}, E.~W., {et~al.} 2010, Astrophysical Journal, 717, 603, \dodoi{10.1088/0004-637X/717/2/603}

\bibitem[{Virtanen {et~al.}(2020)Virtanen, Gommers, Oliphant, Haberland, Reddy, Cournapeau, Burovski, Peterson, Weckesser, Bright, {van der Walt}, Brett, Wilson, Millman, Mayorov, Nelson, Jones, Kern, Larson, Carey, Polat, Feng, Moore, {VanderPlas}, Laxalde, Perktold, Cimrman, Henriksen, Quintero, Harris, Archibald, Ribeiro, Pedregosa, {van Mulbregt}, \& {SciPy 1.0 Contributors}}]{scipy}
Virtanen, P., Gommers, R., Oliphant, T.~E., {et~al.} 2020, Nature Methods, 17, 261, \dodoi{10.1038/s41592-019-0686-2}

\end{thebibliography}
\bibliographystyle{aasjournal}

\end{document}